\documentclass[showkeys,showpacs,superscriptaddress]{revtex4}
\usepackage[caption=false]{subfig}
\usepackage{amsmath}
\usepackage{amssymb}
\usepackage{graphicx}
\usepackage{soul,color}
\usepackage{longtable}
\usepackage{hyperref}
\usepackage[normalem]{ulem}

\usepackage{makecell}
\usepackage{flafter}

\begin{document}

\title{
Non-Abelian Proca-Dirac-Higgs theory: \\particle-like solutions and their energy spectrum
}

\author{
Vladimir Dzhunushaliev
}
\email{v.dzhunushaliev@gmail.com}
\affiliation{
Department of Theoretical and Nuclear Physics,  Al-Farabi Kazakh National University, Almaty 050040, Kazakhstan
}
\affiliation{
Institute of Experimental and Theoretical Physics,  Al-Farabi Kazakh National University, Almaty 050040, Kazakhstan
}
\affiliation{
Institute of Physicotechnical Problems and Material Science of the NAS of the Kyrgyz Republic, 265 a, Chui Street, Bishkek 720071, Kyrgyzstan
}
\affiliation{Institute of Systems Science, Durban University of Technology, 4000 Durban, South Africa}

\author{Vladimir Folomeev}
\email{vfolomeev@mail.ru}
\affiliation{
Institute of Experimental and Theoretical Physics,  Al-Farabi Kazakh National University, Almaty 050040, Kazakhstan
}
\affiliation{
Institute of Physicotechnical Problems and Material Science of the NAS of the Kyrgyz Republic, 265 a, Chui Street, Bishkek 720071, Kyrgyzstan
}

\author{Arislan Makhmudov}
\email{arslan.biz@gmail.com}
\affiliation{Institute of Systems Science, Durban University of Technology, 4000 Durban, South Africa}


\begin{abstract}
We study a system consisting of a non-Abelian SU(2) Proca field interacting with nonlinear scalar (Higgs) and spinor fields.
For such a system, it is shown that particle-like solutions with finite energy do exist.
It is demonstrated that the solutions depend on three free parameters of the system, including the central value of the scalar field $\xi_0$.
For some fixed values of $\xi_0$, we find energy spectra of the solutions. It is shown that for each of the cases under consideration there is
a minimum value of the energy $\Delta=\Delta(\xi_0)$
 (the mass gap $\Delta(\xi_0)$ for a fixed value of $\xi_0$). The behavior of the function
$\Delta(\xi_0)$  is studied for some range of  $\xi_0$.
\end{abstract}

\pacs{
11.90.+t
}

\keywords{
non-Abelian SU(2) Proca theory, nonlinear Dirac equation, Higgs field,  particle-like solutions, energy spectrum, mass gap
}
\date{\today}

\maketitle

\section{Introduction}

By Proca theories one means gauge theories (both Abelian and non-Abelian ones) where the gauge invariance is violated explicitly by introducing a mass term.
The first use of a Proca field was by Yukawa to describe pions. Later on the Proca theory has found applications in various fields of modern theoretical physics.
The use of a Proca field results in the following consequences: a photon may acquire a rest mass;
Einstein-Proca gravity involves a graviton of nonzero rest mass;  Einstein-Proca-Weyl theories can be applied to describe dark matter~\cite{Scipioni:1999mf};
using a real Proca field with a negative mass square, one can describe tachyons~-- particles moving with the velocity greater than the velocity of light~\cite{Tomaschitz:2005fy}.

Additionally, the following researches involving a Proca field may be noted.
Ref.~\cite{Ponglertsakul:2016fxj} studies Einstein-non-Abelian-Proca theory
in an asymptotically anti-de Sitter spacetime with gauge group SU(2). The results obtained describe a gravitating Proca monopole. Refs.~\cite{Brito:2015pxa}
and \cite{Herdeiro:2017fhv} consider stars supported by Proca fields. In Ref.~\cite{Heisenberg:2014rta},  theories of the generalized Proca field are under investigation:
the generalized Proca action for a vector field with derivative self-interactions with only three propagating degrees of freedom is constructed.
The paper~\cite{Allys:2015sht} also deals with the generalized Proca action for an Abelian vector field. The work~\cite{Silenko:2017qgo}
considers  relativistic quantum mechanics of a Proca particle in Riemannian spacetimes. In Ref.~\cite{deFelice:2017paw} the authors have placed
observational constraints on a class of dark energy models within the framework of generalized Proca theories. The purpose of the paper~\cite{Allys:2016kbq}
is to build the first-order terms of the generalized SU(2) Proca theory and to discuss a general form of the complete theory.

In the present paper we study a non-Abelian SU(2) Proca field interacting with nonlinear scalar and spinor fields. The scalar field is described by the
Klein-Gordon equation with the Higgs potential. The spinor field $\psi$ is described by the Dirac equation with a potential term of the form
$\left| \bar \psi \psi \right|^2$. Our purpose here is to obtain particle-like spherically symmetric solutions and to study their energy spectra.
We will show below
that the energy spectrum depends on the parameters $f_2, E$, and $\phi_0$. The parameter  $f_2$ describes
a behaviour of the SU(2) Proca field at the center of the system;
$E/\hbar$ is a frequency of the stationary  spinor field entering the factor $e^{-i E t / \hbar}$; the parameter $\phi_0$ is a central value of the Higgs field $\phi$.
We show that the energy spectrum has a minimum, at least for some values of $\phi_0$, and we argue that this will also take place for any value of $\phi_0$
lying in the range $0< \phi_0 <\infty$. The behaviour of this minimum as $\phi_0 \rightarrow \infty$ is of great interest: if in this limit the minimum is nonzero,
one can say that there is a mass gap $\Delta \neq 0$ in non-Abelian Proca-Dirac-Higgs theory.


If such a mass gap does exist, this would be of great significance. The reason is that in quantum field theory there is a problem to prove the existence
of a mass gap in quantum  chromodynamics
Since this problem is highly nontrivial, one of possible ways for solving it could be a consideration of simpler problems where
\emph{quantum systems} are replaced by \emph{approximate  classical systems}. In this case,
if one could show that in such classical systems a mass gap can occur,  this could be regarded as an indication of the possibility
 of existence of the mass gap in quantum systems. From this point of view, the classical system studied in the present paper and regarded as some approximation to
 realistic quantum systems can be of some interest.

Thus, the purpose of the present paper is to (i) obtain particle-like solutions within a theory with a non-Abelian SU(2) Proca field plus a Higgs scalar field plus a nonlinear Dirac field;
(ii) study energy spectra of these solutions; (iii) search for a minimum of the spectrum (a mass gap); and
(iv) understand the mechanism of the occurrence of a mass gap within the theory under investigation and, on this basis,  to suggest a similar mechanism for QCD.

The paper is organized as follows. In Sec.~\ref{Proca_Dirac_scalar}, we give the field equations describing a system consisting of a non-Abelian SU(2) Proca field interacting with
nonlinear scalar and spinor fields. In Sec.~\ref{QMplusQuarks}, we choose the ansatzs (stationary one for the spinor field and static ones for the Proca and scalar fields)
to solve the equations of Sec.~\ref{Proca_Dirac_scalar}, using which the corresponding complete set of equations is written down. For this set, in Sec.~\ref{QMS} we numerically
solve the equations and find regular solutions describing
particle-like systems in non-Abelian Proca-Dirac-Higgs theory. In Sec.~\ref{MassGap}, we construct the energy spectrum of the solutions obtained and show the existence of a mass
gap for this spectrum. Finally, in Sec.~\ref{concl}, we summarize and discuss the results obtained.

\section{Non-Abelian Proca plus Higgs scalar and nonlinear Dirac fields theory}
\label{Proca_Dirac_scalar}

The Lagrangian describing a system consisting of a non-Abelian SU(2) Proca field $A^a_\mu$ interacting with nonlinear scalar,
$\phi$, and spinor, $\psi$, fields can be taken in the form
\begin{equation}
\begin{split}
	\mathcal L = & - \frac{1}{4} F^a_{\mu \nu} F^{a \mu \nu} -
	\frac{\mu^2}{2} A^a_\mu A^{a \mu} +
	\frac{1}{2} \partial_\mu \phi \partial^\mu \phi +
	\frac{\lambda_1}{2} \phi^2 A^a_\mu A^{a \mu} -
	\frac{\lambda_2}{4} \left( \phi^2 - M^2 \right)^2
\\
	&
	+i \hbar c \bar \psi \gamma^\mu \left(
		\partial_\mu - i \frac{g}{2} \sigma^a A^a_\mu
	\right) \psi +
	\frac{\Lambda}{2} g \hbar c \phi \left( \bar \psi \psi \right)^2 -
	m_f c^2 \bar \psi \psi.
\label{1-05}
\end{split}
\end{equation}
Here $m_f$ is the mass of the spinor field;
$
	D_\mu = \partial_\mu - i \frac{g}{2} \sigma^a
	A^a_\mu
$ is the gauge derivative, where $g$ is the coupling constant and $\sigma^a$ are the SU(2) generators (the Pauli matrices);
$
	F^a_{\mu \nu} = \partial_\mu A^a_\nu - \partial_\nu A^a_\mu +
	g \epsilon_{a b c} A^b_\mu A^c_\nu
$ is the tensor of the Proca field in non-Abelian SU(2) Proca theory, where $\epsilon_{a b c}$ (the completely antisymmetric Levi-Civita symbol)
are the SU(2) structure constants;  $\mu, M, \Lambda$, and $\lambda_{1,2}$ are constants; $\gamma^\mu$ are the Dirac matrices in the standard representation.

The corresponding field equations are as follows:
\begin{eqnarray}
	D_\nu F^{a \mu \nu} - \left(
		\lambda_1 \phi^2 - \mu^2
	\right) A^{a \mu} &=& \frac{g \hbar c}{2}
		\bar \psi \gamma^\mu \sigma^a \psi ,
\label{1-10}\\
	\Box \phi - \lambda_1 A^a_\mu A^{a \mu} \phi -
	\lambda_2 \phi \left( M^2 - \phi^2 \right) &=&
	\frac{\Lambda}{2} g \hbar \left(
		\bar \psi \psi
	\right)^2 ,
\label{1-20}\\
	i \hbar \gamma^\mu \left(
		\partial_\mu  - i \frac{g}{2} \sigma^a
		A^a_\mu
	\right) \psi + \Lambda g \hbar \phi \psi
	\left(
		\bar \psi \psi
	\right) - m_f c \psi &=& 0.
\label{1-30}
\end{eqnarray}

Let us note the following features of this set of equations:
(i) the presence of the term
$\left(\lambda_1 \phi^2 - \mu^2\right)$ in Eq.~\eqref{1-10} will lead to exponential damping of the  SU(2) Proca field, and this is a distinctive feature of the Proca monopole
compared with the 't~Hooft-Polyakov monopole;
(ii) the Higgs field $\phi$ is topologically trivial, and this is also the distinction of principle compared with the 't~Hooft-Polyakov monopole;
(iii) the nonlinear spinor field $\psi$ allows the existence of particle-like solutions even in the absence of the fields $A^a_\mu$ and $\phi$ (see Refs.~\cite{Finkelstein:1951zz,Finkelstein:1956});
(iv) the system supported by the fields  $A^a_\mu$ and $\phi$ also allows the existence of particle-like solutions~-- a Proca monopole;
(v) particle-like solutions exist only for some special choices of the constants $\mu, M$, and $u_1$ [for the definition of parameter $u_1$, see Eq.~\eqref{3-c-70}];
(vi)~in the absence of the fields $A^a_\mu$ and $\phi$, the nonlinear Dirac equation  \eqref{1-30}  describes a system with a  mass gap (see  Refs.~\cite{Finkelstein:1951zz,Finkelstein:1956}),
in contrast to Eqs.~\eqref{1-10} and \eqref{1-20} which yield a system whose energy spectrum has no mass gap.

To obtain particle-like solutions, the above equations will be solved numerically since apparently it is impossible to find their analytical solution.
As will be shown below, these particle-like solutions have the following interesting feature:
For each central value of the scalar field $\phi_0$ the plane $\{f_2, E\}$ [where $E/\hbar$ is the spinor frequency, see Eq.~\eqref{2-40}, and
$f_2$ is a free parameter associated with the Proca field, see Eq.~\eqref{3-c-50}]
is divided into two regions: in one of them,
particle-like solutions do exist, and in the other one they are absent. This means that there is some curve $\gamma$ dividing these two regions. Near this curve
there is a complicated interaction between the fields $A^a_\mu, \phi$, and $\psi$, which possibly results in an infinite energy of the particle-like solution on the curve $\gamma$.

\section{Equations for a Proca monopole interacting with nonlinear Higgs and Dirac fields}
\label{QMplusQuarks}

We seek particle-like solutions describing objects supported by a radial magnetic field (a Proca monopole or a Proca hedgehog) and a nonlinear spinor field.
For this purpose, we employ the standard SU(2) monopole ansatz \eqref{2-20} and the ansatz \eqref{2-40} for the spinor field, each row of which
describes a spinor for the ground state of an electron in a hydrogen atom; i.e., we seek a solution of Eqs.~\eqref{1-10}-\eqref{1-30} in the following form:
\begin{eqnarray}
	A^a_i &=& - \frac{1}{g} \left[ 1 - f(r) \right]
	\begin{pmatrix}
		 0 & \phantom{-}\sin \varphi &  \sin \theta \cos \theta \cos \varphi \\
		 0 & \cos \varphi &  - \sin \theta \cos \theta \sin \varphi \\
		 0 & 0 & - \sin^2 \theta
	\end{pmatrix} , \quad
	i = r, \theta, \varphi  \text{ (in polar coordinates)},
\label{2-20}\\
	A^a_t &=& 0 ,
\label{2-13}\\
	\phi &=& \frac{\xi(r)}{g},
\label{2-30}\\
	\psi^T &=& \frac{e^{-i \frac{E t}{\hbar}}}{g r \sqrt{2}} \begin{Bmatrix}
		\begin{pmatrix}
			0 \\ - u \\
		\end{pmatrix},
		\begin{pmatrix}
			u \\ 0 \\
		\end{pmatrix},
		\begin{pmatrix}
			i v \sin \theta e^{- i \varphi} \\ - i v \cos \theta \\
		\end{pmatrix},
		\begin{pmatrix}
			- i v \cos \theta \\ - i v \sin \theta e^{i \varphi} \\
		\end{pmatrix}
	\end{Bmatrix},
\label{2-40}
\end{eqnarray}
where $E/\hbar$ is the spinor frequency and the color index $a=1,2,3$.  The functions $u$ and $v$ depend on the radial coordinate $r$ only. The ansatz~\eqref{2-40} is taken from Refs.~\cite{Li:1982gf,Li:1985gf}.
After substituting the expressions \eqref{2-20}-\eqref{2-40} into the Lagrangian \eqref{1-05}, we have
\begin{equation}
\begin{split}
	\mathcal{\tilde{L}_{\text{eff}}} \equiv & \frac{\mathcal{L_{\text{eff}}}}{\hbar c/r_0^4} =
 	\frac{1}{\tilde g^2} \left\lbrace -
	\left(
		\frac{{f'}^2}{ x^2} +
		\frac{\left( f^2 - 1 \right)^2}{2 x^4} -
		\tilde \mu^2 \frac{\left( f - 1 \right)^2}{x^2}
	\right) -
	2 m^2 \left[
		\tilde \xi^{\prime 2} +
		\tilde\lambda_1 \frac{\left( f - 1 \right)^2}{2 x^2} \tilde \xi^2 +
		\frac{\tilde \lambda}{2}\left(
		\tilde \xi^2 - \tilde M^2
		\right)^2
	\right]
	\right\rbrace
\\
	&	
	+ \frac{1}{ x^2} \left[
		- \tilde u \tilde v' + \tilde u' \tilde v -
		2 f \frac{\tilde u \tilde v}{x} +
		m^2 \frac{\tilde\Lambda}{2} \tilde \xi
		\frac{\left(\tilde u^2 - \tilde v^2 \right)^2}{x^2} -
		\tilde m_f \left(\tilde  u^2 - \tilde v^2 \right) +
		\tilde E \left(\tilde u^2 + \tilde  v^2 \right)
		\right] .
\label{2-90}
\end{split}
\end{equation}
Here, for convenience of performing numerical calculations, we have introduced the following dimensionless variables:
$\tilde g^2 = g^2 \hbar c$ is the dimensionless coupling constant for the SU(2) Proca gauge field; $x = r/r_0$, where $r_0$ is a constant corresponding to the characteristic size of the system under consideration;
$\tilde u=\sqrt{r_0}u/g$,
$
\tilde v = \sqrt{r_0}v/g,
\tilde \mu = r_0 \mu,
\tilde \xi = r_0 g \phi / 2m,
\tilde M = g r_0 M / 2m,
\tilde \lambda_1 = 4 \lambda_1/g^2,
\tilde \lambda = 4 m^2 \lambda_2 /g^2,
\tilde m_f = r_0 m_f c/\hbar,
\tilde E = r_0 E/(\hbar c),
\tilde \Lambda = 2 \Lambda/(m r_0^3)
$;
$m$ is a free parameter introduced for convenience;
the prime denotes differentiation with respect to  $x$. 
The parameter $r_0$ must depend only on constants of a theory; therefore it can be taken, for instance, in the form
 $r_0 = \alpha \hbar/(m_f c)$, where $\alpha$ is a constant.

Equations for the unknown functions $f(x), \xi(x), u(x)$, and $v(x)$ can be obtained either by substituting Eqs.~\eqref{2-20}-\eqref{2-40}
into the field equations  \eqref{1-10}-\eqref{1-30} or by varying the Lagrangian \eqref{2-90} with respect to the corresponding functions $f(x), \xi(x), u(x)$, and $v(x)$.
However, in the latter case one should take into account that the nonlinear term
$
	\left(
		\bar \psi \psi
	\right)^2 =
	\frac{\left(\tilde u^2 - \tilde v^2 \right)^2}{x^4}
$
must be written in the form
$
\frac{\left(\tilde u^2_\psi - \tilde v^2_\psi \right)
\left(\tilde u^2_{\bar\psi} - \tilde v^2_{\bar\psi} \right)}{x^4},
$
where $(\tilde u, \tilde v)_{\bar\psi}$ are taken from $\bar{\psi}$ and
$(u, v)_{\psi}$ -- from $\psi$. Therefore the corresponding Dirac equation is obtained by varying \eqref{2-90} with respect to $(\tilde u, \tilde v)_{\bar\psi}$, and then one must take
$
	(\tilde u, \tilde v)_{\bar\psi} = (\tilde u, \tilde v)_{\psi} =
	(\tilde u, \tilde v)
$. Thus we obtain the following equations:
\begin{eqnarray}
	- f^{\prime \prime} + \frac{f \left( f^2 - 1 \right) }{x^2} -
	m^2 \left( 1 - f \right) \tilde \xi^2 +
	\tilde g^2\,\frac{\tilde u \tilde v}{x} &=& - \tilde{\mu}^2 \left( 1 - f \right) ,
\label{2-50}\\
	\tilde \xi^{\prime \prime} + \frac{2}{x} \tilde \xi^\prime &=&
	\tilde \xi \left[
		\frac{\left( 1 - f \right)^2}{2 x^2}  +
		\tilde{\lambda} \left(
			\tilde \xi^2 - \tilde{M}^2
		\right)
	\right] -
	\frac{\tilde g^2\tilde\Lambda}{8}
	\frac{\left( \tilde u^2 - \tilde  v^2 \right)^2}{x^4} ,
\label{2-60}\\
	\tilde v' + \frac{f \tilde v}{x} &=& \tilde u \left(
		- \tilde m_f + \tilde E +
		m^2 \tilde \Lambda   \frac{\tilde u^2 - \tilde v^2}{x^2} \tilde \xi
	\right) ,
\label{2-70}\\
	\tilde u' - \frac{f \tilde u}{x} &=& \tilde v \left(
		- \tilde m_f - \tilde E +
				m^2 \tilde \Lambda   \frac{\tilde u^2 - \tilde v^2}{x^2} \tilde \xi
	\right).
\label{2-80}
\end{eqnarray}
For convenience of performing numerical calculations,
we have taken $\tilde{\lambda}_1 = 1$. Eqs.~\eqref{2-50} and \eqref{2-60} describe the Proca monopole with the sources appearing due to the presence
of the Dirac and Higgs fields. We emphasize that these equations differ considerably from those describing the 't~Hooft-Polyakov monopole. Eqs.~\eqref{2-70} and \eqref{2-80}
have been studied in a simplified form for the case of $f = 1$ and $\tilde \xi = \mathrm{const}$ in  Refs.~\cite{Finkelstein:1951zz,Finkelstein:1956},
where it was shown that this set of equations has particle-like solutions whose energy spectrum possesses a mass gap
(for brevity, we will refer to such solution as a spinball).

Next, by definition, the energy density of the spinor field is
\begin{equation}
	\epsilon_s =i\hbar \bar\zeta \gamma^0 \dot\zeta - L_{D, \text{eff}},
\label{2-100}
\end{equation}
where the dot denotes differentiation with respect to time. The Lagrangian of the Dirac field $L_{D, \text{eff}}$ appearing here is given by the expression from \eqref{2-90},
\begin{equation}
\begin{split}
	\tilde L_{D, \text{eff}} = &
	\frac{1}{ x^2} \left[
		- \tilde u \tilde v' + \tilde u' \tilde v -
		2 f \frac{\tilde u \tilde v}{x} +
		m^2 \frac{\tilde\Lambda}{2} \tilde \xi
		\frac{\left(\tilde u^2 - \tilde v^2 \right)^2}{x^2} -
		\tilde m_f \left(\tilde  u^2 - \tilde v^2 \right) +
		\tilde E \left(\tilde u^2 + \tilde  v^2 \right)
		\right]
\\
	&
	= - m^2 \frac{\tilde\Lambda}{2 } \tilde \xi
	\frac{\left(\tilde u^2 - \tilde v^2 \right)^2}{x^4},
\end{split}
\label{2-110}
\end{equation}
which is obtained using Eqs.~\eqref{2-70} and \eqref{2-80}. Then, using the ansatz \eqref{2-40}, the energy density of the spinor field \eqref{2-100}
can be found in the following dimensionless form:
\begin{equation}
	\tilde \epsilon_s = \tilde E \frac{\tilde u^2 + \tilde v^2}{x^2} +
	m^2 \frac{\tilde\Lambda}{2 } \tilde \xi
	\frac{\left(\tilde u^2 - \tilde v^2 \right)^2}{x^4} .
\label{2-120}
\end{equation}
As a result, we get the following total energy density of the particle-like solution:
\begin{equation}
	\tilde \epsilon =
 	\tilde{\epsilon}_{\text{Pm}} + \tilde \epsilon_s ,
\label{2-130-b}
\end{equation}
where
\begin{equation}
	\tilde{\epsilon}_{\text{Pm}} = \frac{1}{\tilde g^2} \left\lbrace
		\left[
			\frac{{f'}^2}{ x^2} +
			\frac{\left( f^2 - 1 \right)^2}{2 x^4} -
			\tilde \mu^2 \frac{\left( f - 1 \right)^2}{x^2}
		\right] +
		2 m^2 \left[
			\tilde \xi^{\prime 2} +
			\frac{\left( f - 1 \right)^2}{2 x^2} \tilde \xi^2 +
			\frac{\tilde \lambda}{2}\left(
			\tilde \xi^2 - \tilde M^2
			\right)^2
		\right]
		\right\rbrace
\label{2-130-c}
\end{equation}
is the energy of the Proca monopole. The formula \eqref{2-130-b} is remarkable because the total energy of the particle-like solution is a sum
of energies of the Proca monopole and of the spinball,
despite the strong interaction between the fields $f$ and  $\xi$, which make up the Proca monopole, and the fields
 $u$ and $v$ supporting the spinball.

\section{Particle-like solutions: a Proca monopole plus a spinball}
\label{QMS}

This section is devoted to studying particle-like solutions of Eqs.~\eqref{2-50}-\eqref{2-80}.
Since apparently there is no analytical solution of this set of equations, we seek a numerical solution. Because of the presence
of terms containing $x$ in the denominators of Eqs.~\eqref{2-50}-\eqref{2-80}, to perform numerical computations, we assign boundary conditions near the origin
 $x = 0$  where solutions are sought in the form of the Taylor series
 \begin{eqnarray}
	f &=& 1 + \frac{f_2}{2} x^2 + \ldots ,
\label{3-c-50}\\
	\tilde\xi &=& \tilde\xi_0 + \frac{\tilde\xi_2}{2} x^2 + \ldots , \quad\quad \text{where} \quad \tilde\xi_2 = - \frac{\tilde g^2 \tilde \Lambda \tilde u_1^4}{24} -
	\frac{\tilde \lambda \tilde \xi_0}{3} \left(
		\tilde \xi_0^2 - \tilde M^2
	\right) ,
\label{3-c-60}\\
	\tilde u &=& \tilde u_1 x + \frac{\tilde u_3}{3!} x^3 + \ldots ,
\label{3-c-70}\\
	\tilde v &=& \frac{\tilde v_2}{2} x^2 + \frac{\tilde v_4}{4!} x^4 + \ldots , \quad \text{where} \quad
	\tilde v_2 = \frac{2}{3} \tilde u_1 \left(
		\tilde E-\tilde m_f  + m^2 \tilde\Lambda \tilde\xi_0  \tilde u_1^2
	\right) .
\label{3-c-80}
\end{eqnarray}
The expansion coefficients $f_2, \tilde \xi_0,$ and $\tilde u_1$ appearing here are free parameters whose values cannot be found from Eqs.~\eqref{2-50}-\eqref{2-80}.

\begin{figure}[!htb]
	\subfloat[The function $\tilde u(x)/x$.
	\label{u_vs_x_E}]{
  	\includegraphics[width=0.45\linewidth]{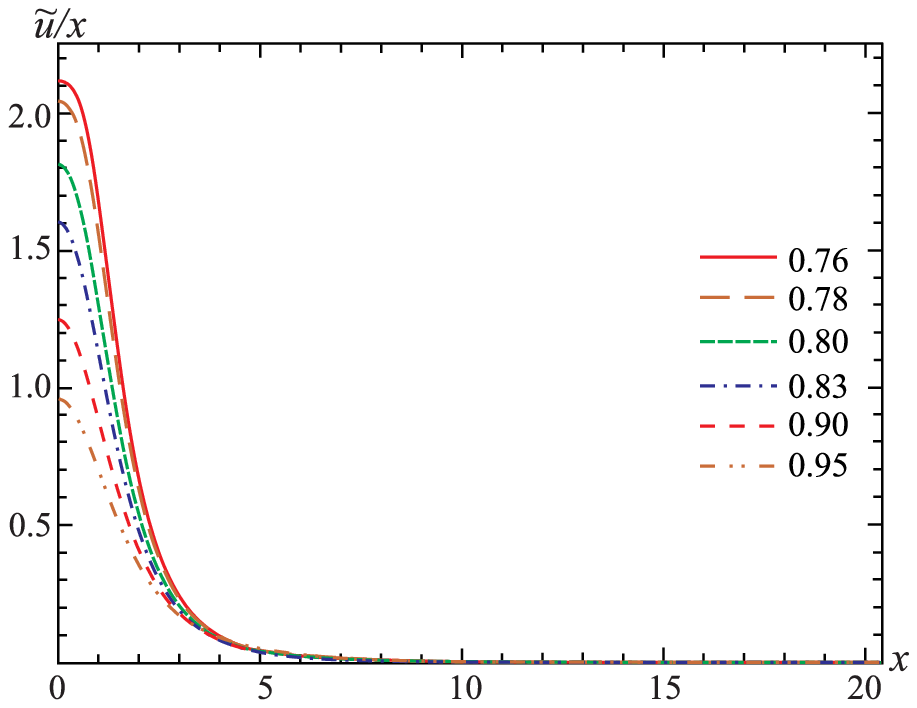}%
	}\hfill
	\subfloat[The function $\tilde v(x)/x$.
	\label{v_vs_x_E}]{%
  	\includegraphics[width=0.45\linewidth]{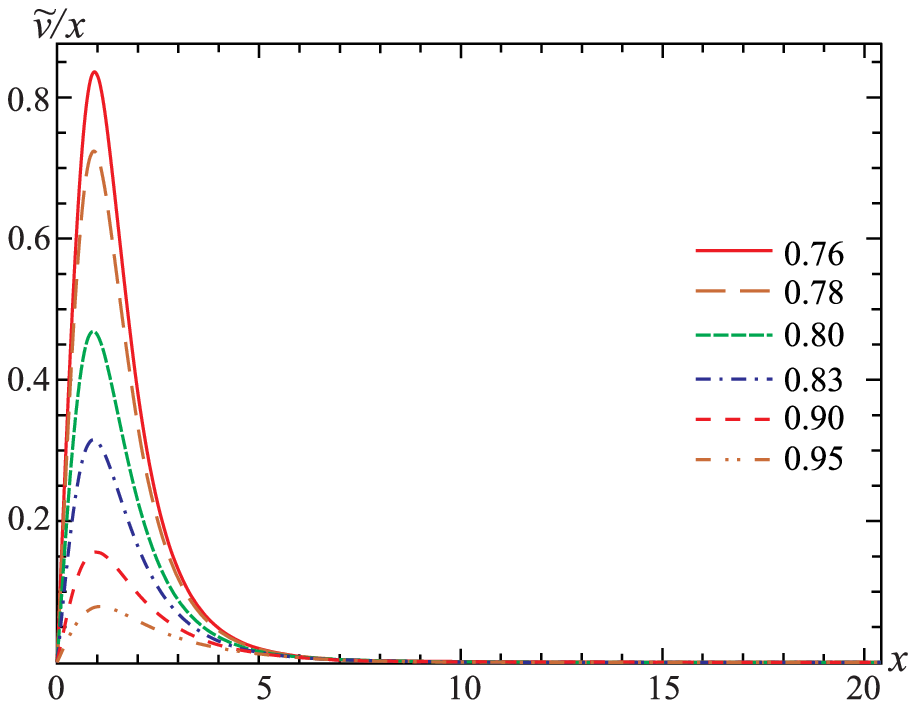}%
	}\hfill
	\subfloat[The function $f(x)$.
	\label{f_vs_x_E}]{
  	\includegraphics[width=0.45\linewidth]{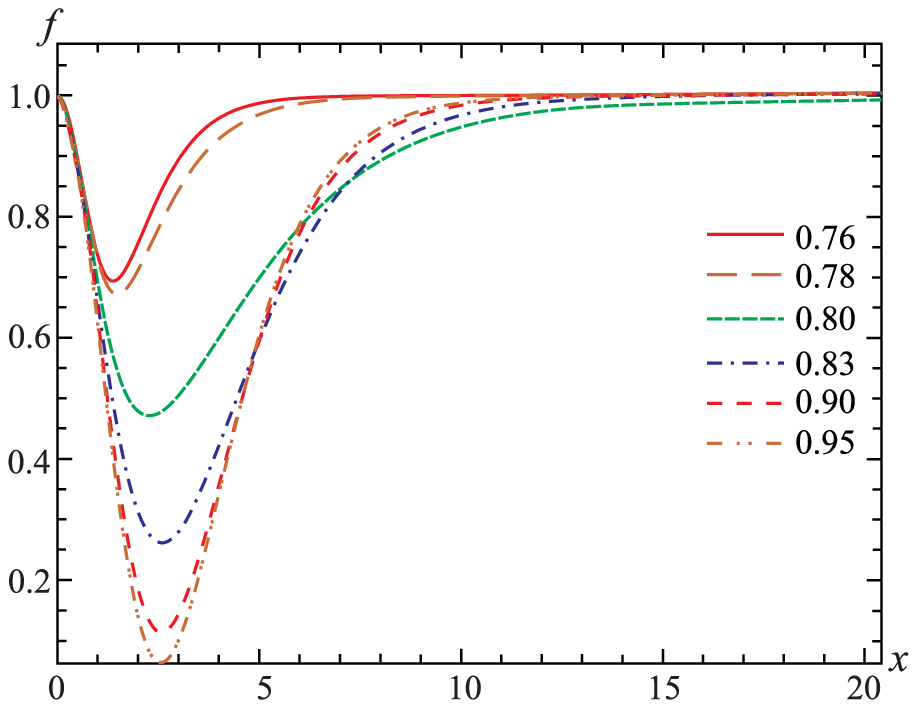}%
	}\hfill
	\subfloat[The function $\tilde \xi(x) - \tilde \xi_0$.
	\label{xi_vs_x_E}]{%
  	\includegraphics[width=0.45\linewidth]{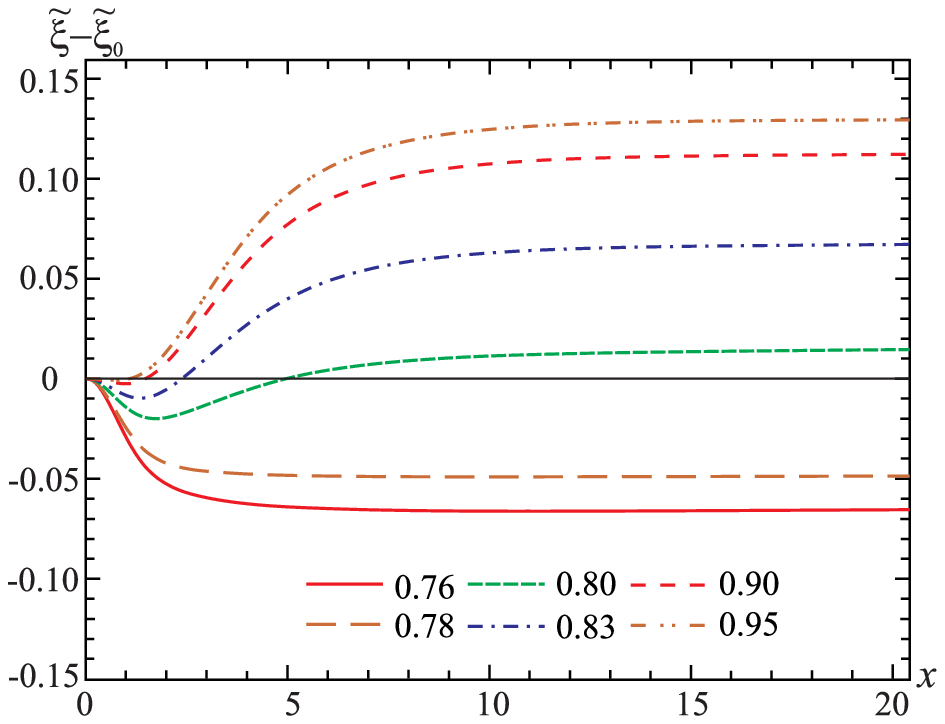}%
	}
\caption{Families of the particle-like solutions for $\tilde \xi_0 = 0.5$ and
	$f_2 = - 	0.95$ for the Proca-monopole-plus-spinball system with
	$\tilde \Lambda = 1/9$, $\tilde m_f = 1$, $m = 3$, $\tilde g = 1$ , $\tilde \lambda = 0.1$ for different values of the parameter $\tilde E$ (designated by the numbers near the curves).
}
\label{uvfxiVs_x_and_E}
\end{figure}

\begin{figure}[!htb]
	\subfloat[The function $\tilde u(x)/x$.
	\label{uVs_x_f2}]{
  	\includegraphics[width=0.45\linewidth]{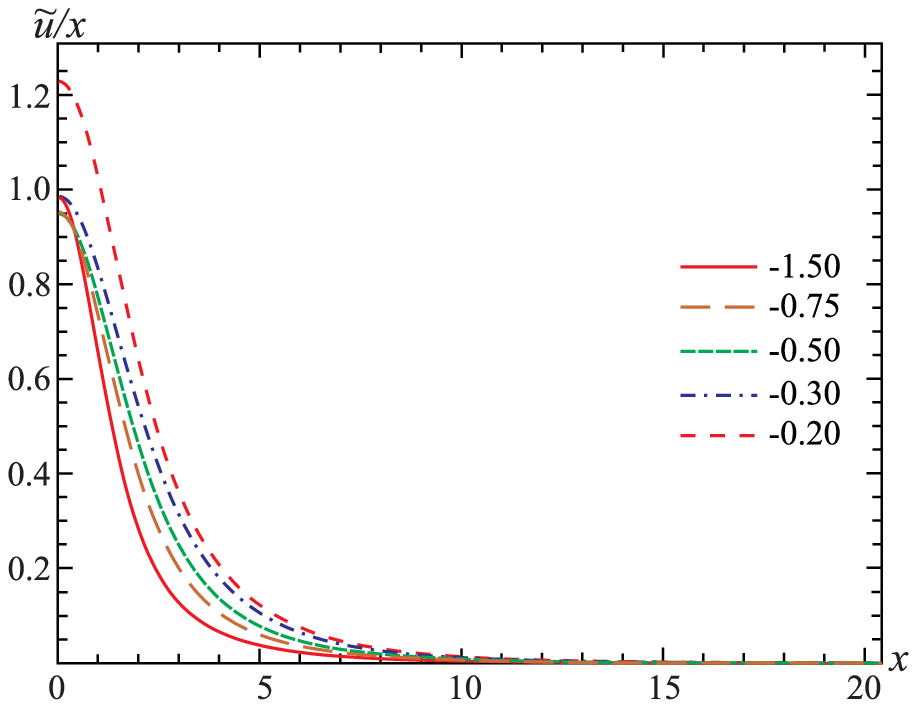}%
	}\hfill
	\subfloat[The function $\tilde v(x)/x$.
	\label{vVs_x_f2}]{%
  	\includegraphics[width=0.45\linewidth]{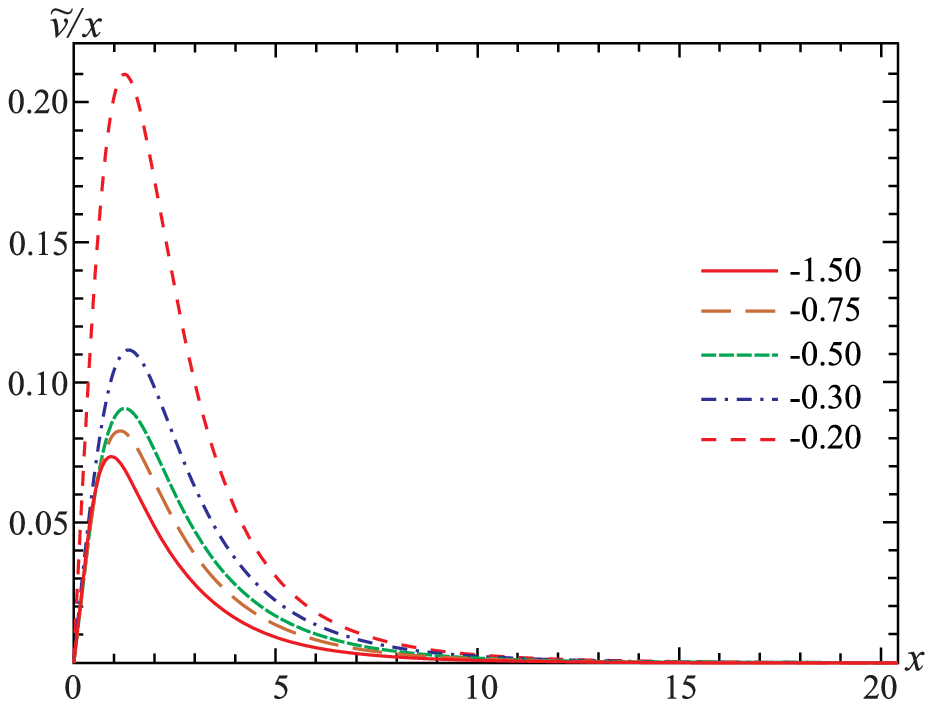}%
	}\hfill
	\subfloat[The function $f(x)$.
	\label{fVs_x_f2}]{
  	\includegraphics[width=0.45\linewidth]{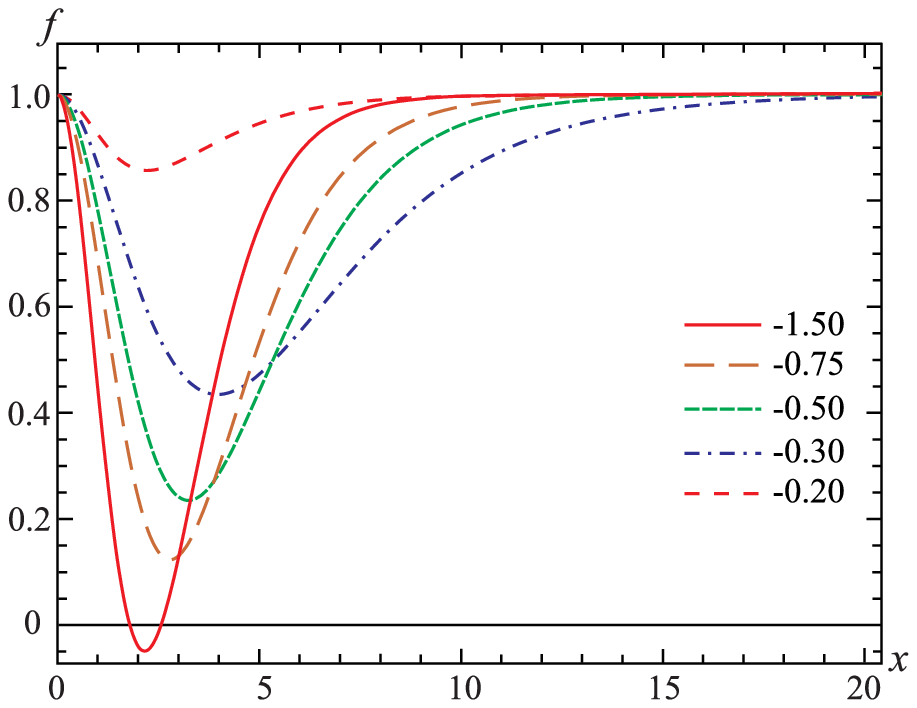}%
	}\hfill
	\subfloat[The function $\tilde \xi(x) - \tilde \xi_0$.
	\label{xiVs_x_f2}]{%
  	\includegraphics[width=0.45\linewidth]{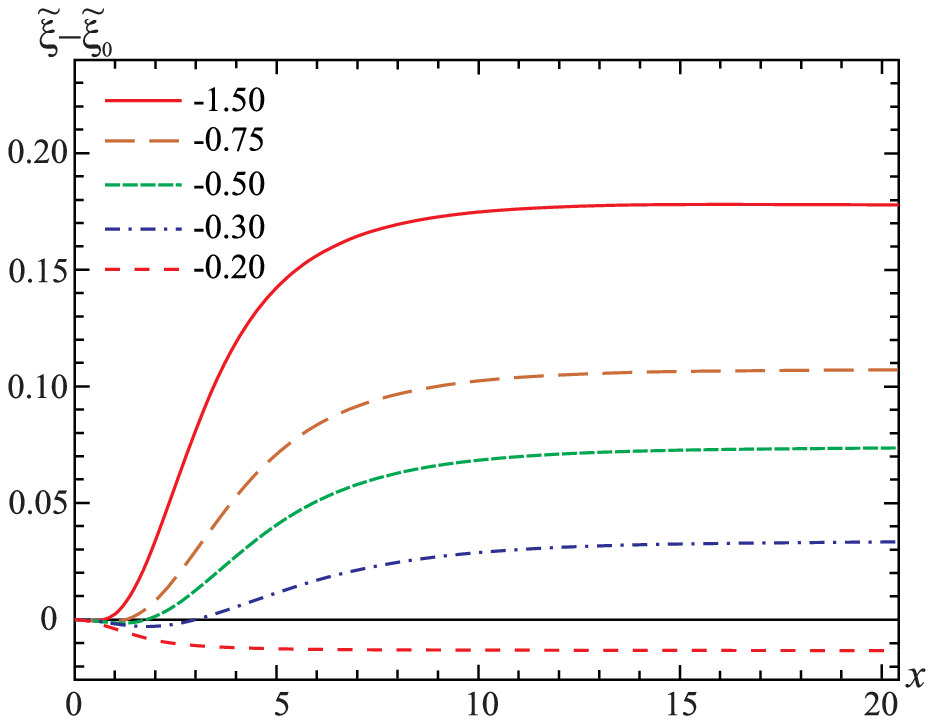}%
	}
\caption{Families of the particle-like solutions for $\tilde \xi_0 = 0.5$ and $\tilde E = 0.95$ for the Proca-monopole-plus-spinball system with
	$\tilde \Lambda = 1/9$, $\tilde m_f = 1$, $m = 3$, $\tilde g = 1$, $\tilde \lambda = 0.1$  for different values of the parameter $f_2$ (designated by the numbers near the curves).
}
\label{uvfxiVs_x_f2}
\end{figure}

\begin{figure}[!htb]
	\subfloat[The function $\tilde u(x)/x$.
	\label{uVs_x_xi0}]{
  	\includegraphics[width=0.45\linewidth]{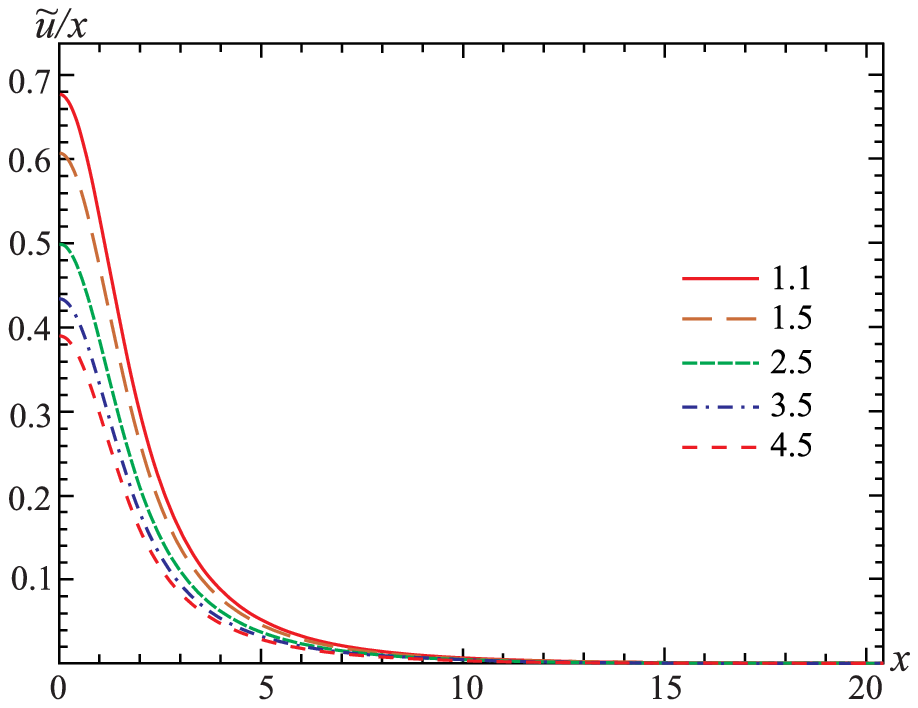}%
	}\hfill
	\subfloat[The function $\tilde v(x)/x$.
	\label{vVs_x_xi0}]{%
  	\includegraphics[width=0.45\linewidth]{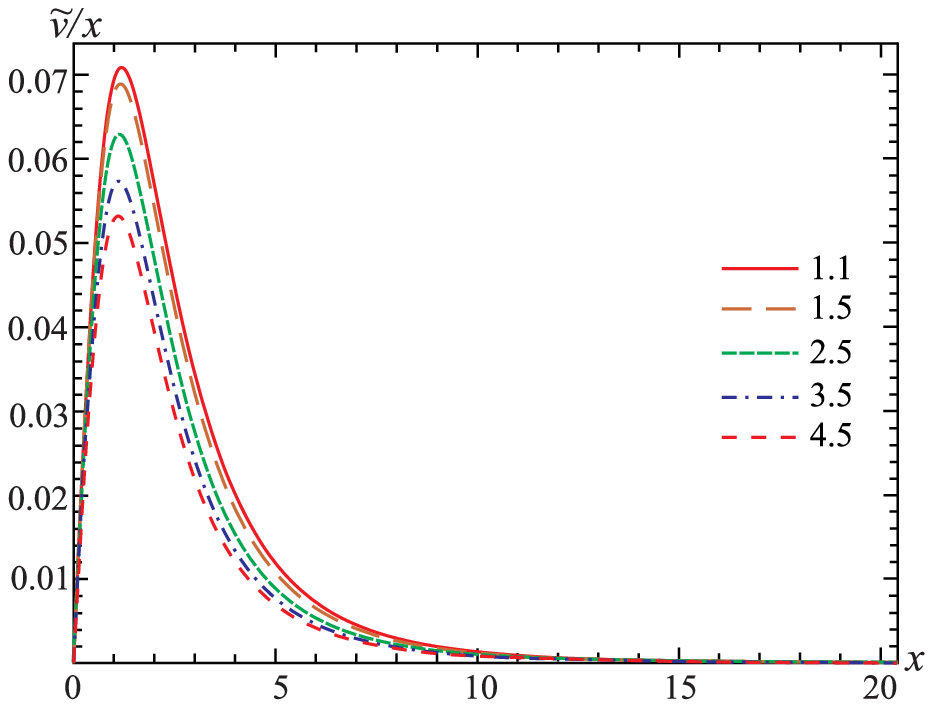}%
	}\hfill
	\subfloat[The function $f(x)$.
	\label{fVs_x_xi0}]{
  	\includegraphics[width=0.45\linewidth]{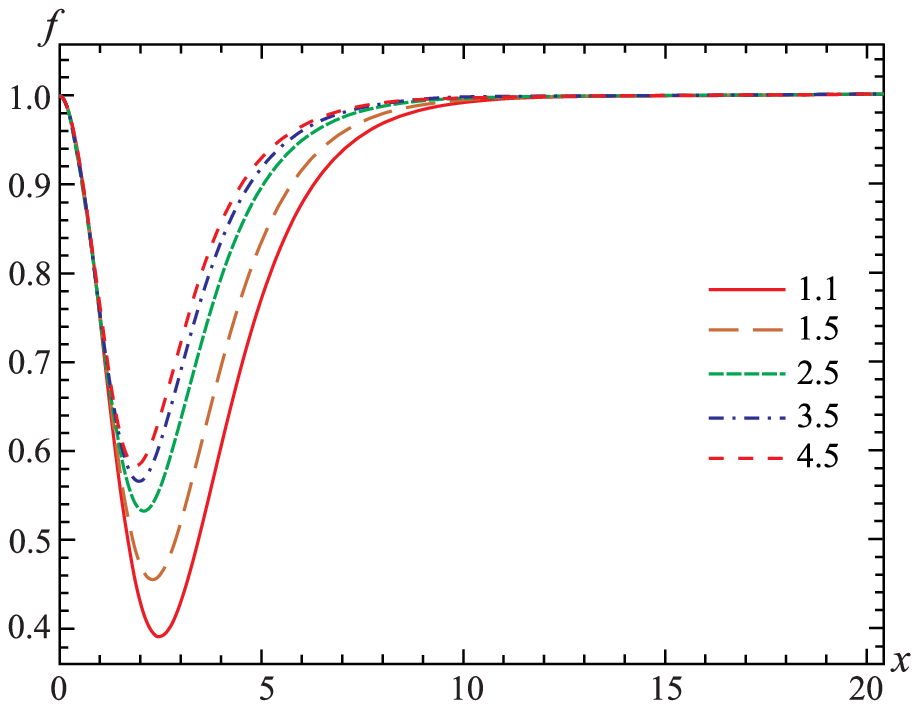}%
	}\hfill
	\subfloat[The function $\tilde \xi(x) - \tilde \xi_0$.
	\label{xiVs_x_xi0}]{%
  	\includegraphics[width=0.45\linewidth]{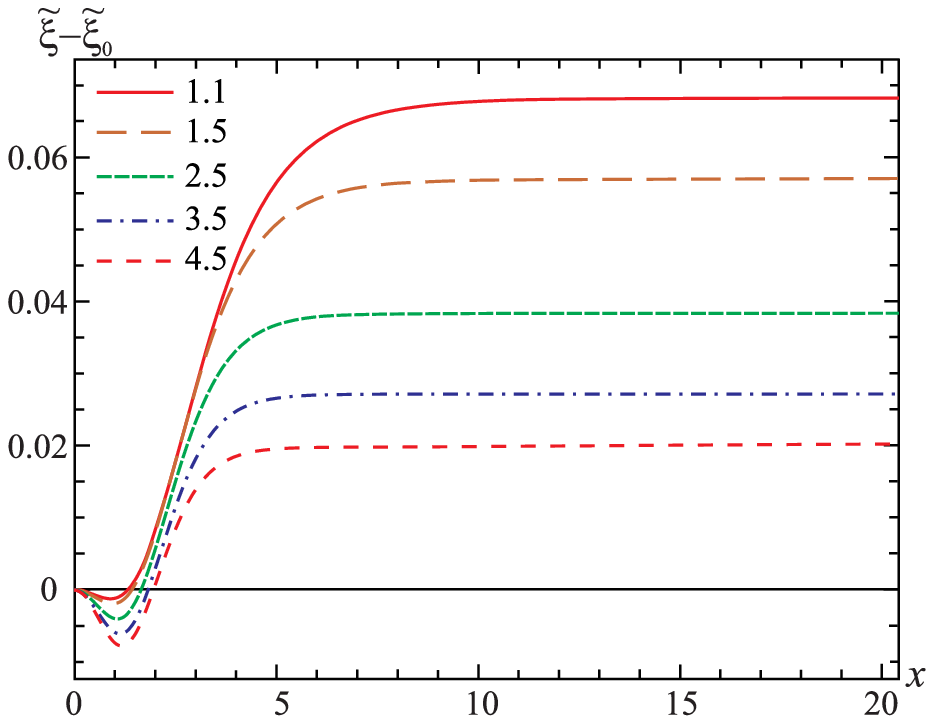}%
	}
\caption{
	Families of the particle-like solutions for $f_2 = - 0.6$ and $\tilde E = 0.96$ for the Proca-monopole-plus-spinball system with
	$\tilde \Lambda = 1/9$, $\tilde m_f = 1$, $m = 3$, $\tilde g = 1$, $\tilde \lambda = 0.1$ for different values of the parameter $\tilde \xi_0$ (designated by the numbers near the curves).
}
\label{uvfxiVs_x_xi0}
\end{figure}

\begin{figure}[!htb]
	\subfloat[The dependence $\tilde M(f_2, \tilde E)$.
	\label{M_vs_E_xi}]{
  	\includegraphics[width=0.48\linewidth]{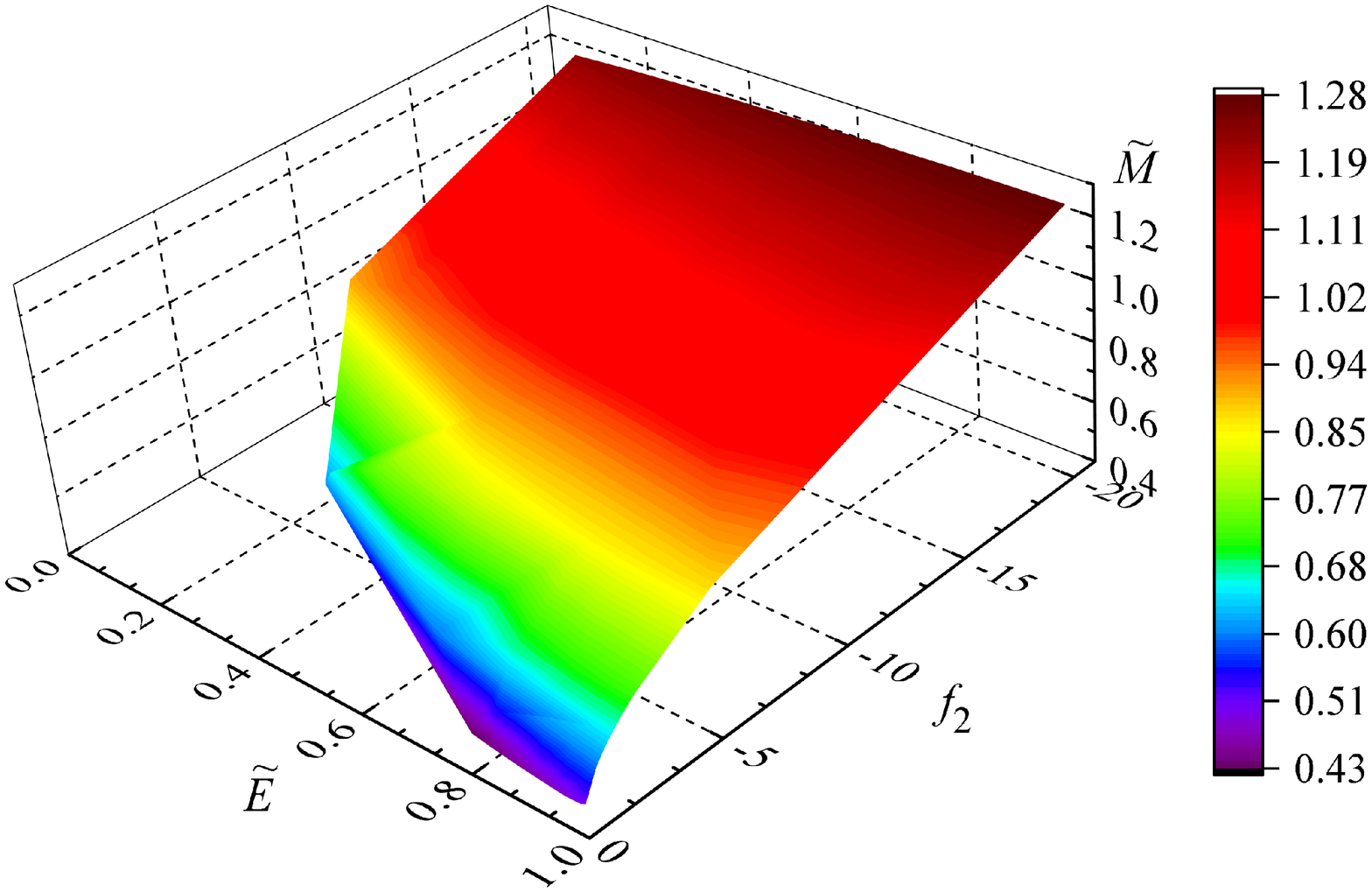}
	}\hfill
	\subfloat[The dependence $\tilde \mu(f_2, \tilde E)$.
	\label{mu_vs_E_xi}]{
  	\includegraphics[width=0.48\linewidth]{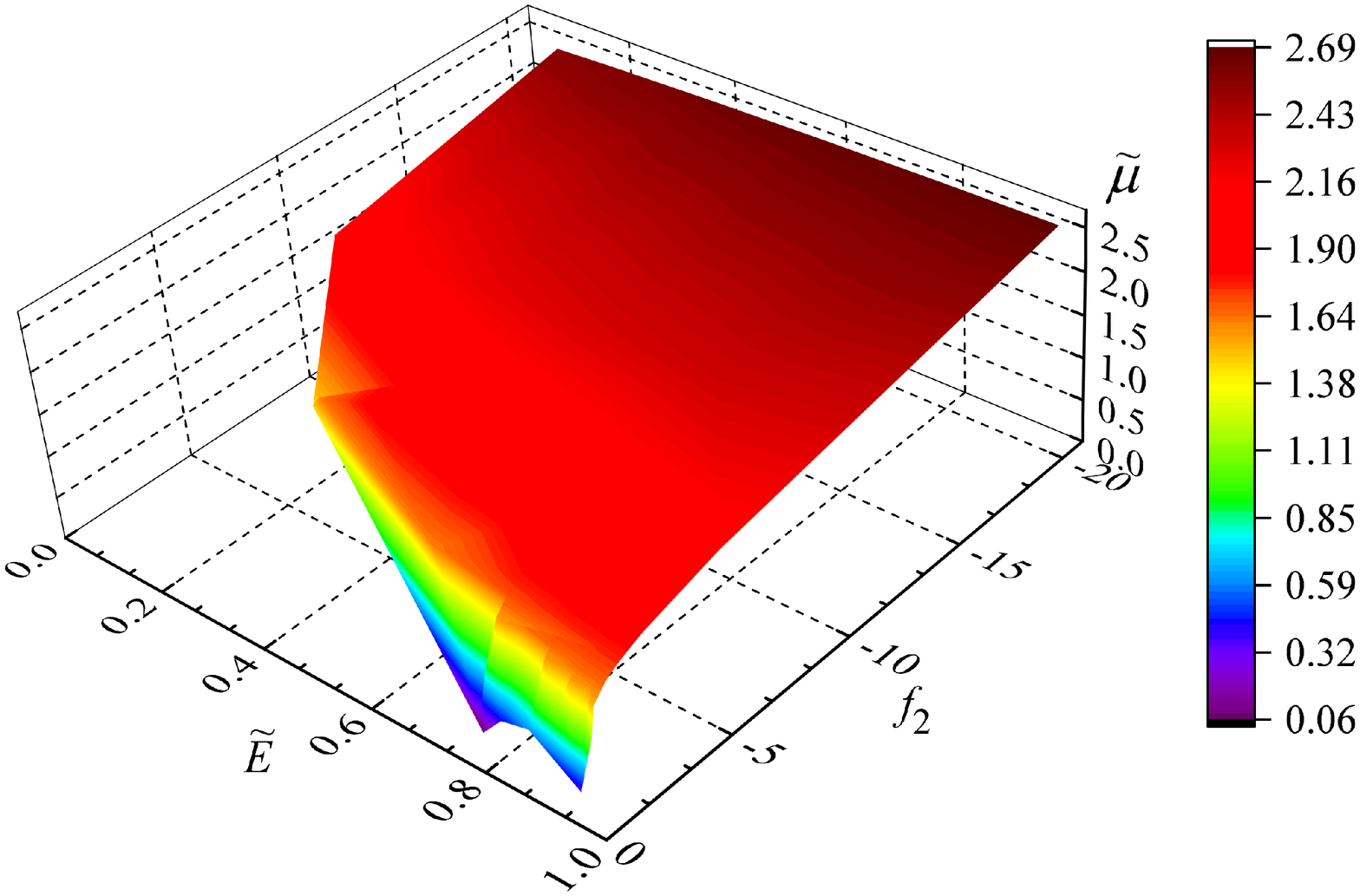}
	}\hfill
\vspace{.5cm}	
\subfloat[The dependence $\tilde{u}_1(f_2, \tilde E)$.
	\label{u1_vs_E_xi}]{
  	\includegraphics[width=0.48\linewidth]{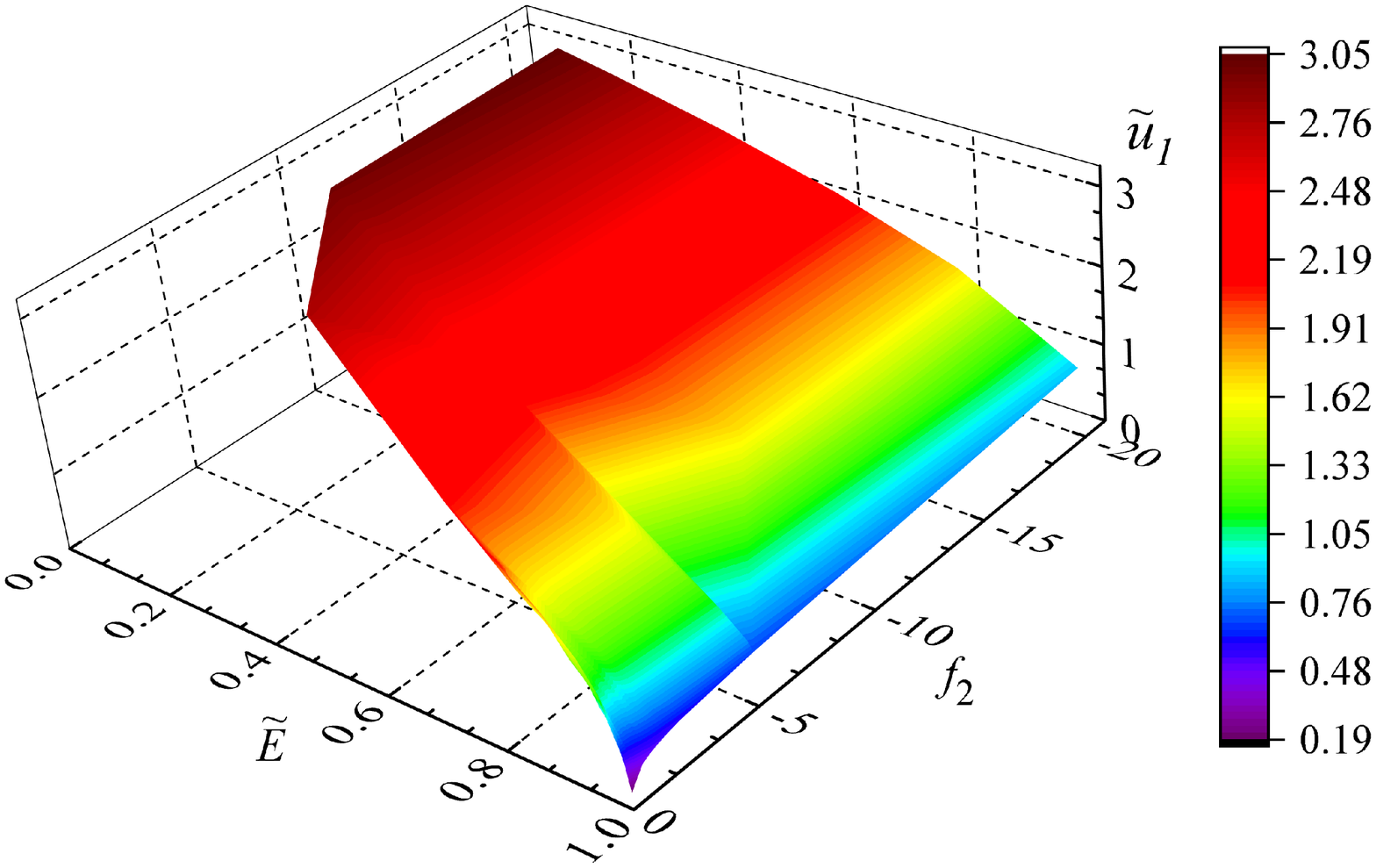}
	}
\caption{The eigenvalues $\tilde M, \tilde \mu$ and $\tilde{u}_1$ as functions of the parameters
 $f_2$ and $\tilde E$ for $\tilde \xi = 0.5$ for the Proca-monopole-plus-spinball system with
	$\tilde \Lambda = 1/9$, $\tilde m_f = 1$, $m = 3$, $\tilde g = 1$, $\tilde \lambda = 0.1$.
}
\label{M_mu_u1_vs_E}
\end{figure}

The equations~\eqref{2-50}-\eqref{2-80} are solved numerically as a nonlinear problem for the eigenvalues $\tilde \mu, \tilde M,$ and $\tilde u_1$
and for the eigenfunctions $f, \tilde\xi, \tilde v,$ and $\tilde u$. Fig.~\ref{uvfxiVs_x_and_E} depicts a typical behaviour
of the solutions for fixed values of $\tilde \xi_0$ and $f_2$,  Fig.~\ref{uvfxiVs_x_f2}~-- for fixed values of $\tilde \xi_0$ and $\tilde E$, and Fig.~\ref{uvfxiVs_x_xi0}~--  for fixed values of $f_2$ and $\tilde E$.
Fig.~\ref{M_mu_u1_vs_E} shows the  eigenvalues
 $\tilde \mu, \tilde M$, and $\tilde u_1$ as functions of the parameters $f_2, \tilde E$.

The asymptotic behaviour of the solutions as $x\to \infty$ is
\begin{align}
\label{3-c-90}
\begin{split}
	f(x) &\approx 1 - f_\infty e^{- x \sqrt{m^2 \tilde{M}^2 - \tilde\mu^2}} -
	\frac{g^2 u_\infty v_\infty}
	{4 \tilde m_f^2 - 4 \tilde E^2 + m^2 \tilde{M}^2 - \tilde\mu^2}
	\frac{e^{- 2 x \sqrt{\tilde m_f^2 - \tilde E^2}}}{x},
	\quad
	\tilde\xi (x) \approx \tilde M - \delta \tilde \xi ,
\\
	\tilde u(x) &\approx \tilde u_\infty
	e^{- x \sqrt{\tilde m_f^2 - \tilde E^2}} ,
	\quad
	\tilde v(x) \approx
	\tilde v_\infty e^{- x \sqrt{\tilde m_f^2 - \tilde E^2}},
\end{split}
\end{align}
where $f_\infty, \tilde\xi_\infty, \tilde v_\infty$,  and $\tilde u_\infty$ are integration constants and the function $\delta \tilde \xi \ll \tilde M$ satisfies the asymptotic equation
\begin{equation}
	\delta \tilde \xi^{\prime \prime} +
	\frac{2}{x} \delta \tilde \xi^{\prime} =
	 - \tilde M \left(
		\frac{f_\infty^2 e^{- 2 x \sqrt{m^2 \tilde{M}^2 - \tilde\mu^2}}}{2 x^2} -
		2 \tilde \lambda \tilde M \delta \tilde \xi
	\right) +
	\frac{\tilde{g}^2 \tilde{\Lambda}}{8} \frac{
		\left(
			\tilde u_\infty^2 - \tilde v_\infty^2
		\right)^2
	}{x^4} e^{- 4 x \sqrt{\tilde m_f^2 - \tilde E^2}} .
\label{3-c-115}
\end{equation}
Its solution can be found in the form
\begin{equation}
	\delta \xi =
	\xi_\infty \frac{e^{- x \sqrt{2 \tilde \lambda \tilde M^2}}}{x} -
	\frac{\tilde M f^2_\infty}{8 \left( m^2 \tilde M^2 - \tilde \mu^2 \right) }
	\frac{e^{- 2 x \sqrt{m^2 \tilde M^2 - \tilde \mu^2}}}{x^2 } +
	\frac{\tilde g^2 \tilde \Lambda
		\left(
			\tilde u_\infty^2 - \tilde v_\infty^2
		\right)^2}{128 \left( \tilde m_f^2 - \tilde E^2 \right) }
		\frac{e^{- 4 x \sqrt{\tilde m_f^2 - \tilde E^2}}}{x^4 } .
\label{3_c_120}
\end{equation}

\begin{figure}[h]
	\subfloat[ $\tilde \xi_0=0.5, \tilde E = 0.95$.
	\label{enVs_x_f2}]{
  	\includegraphics[width=0.45\linewidth]{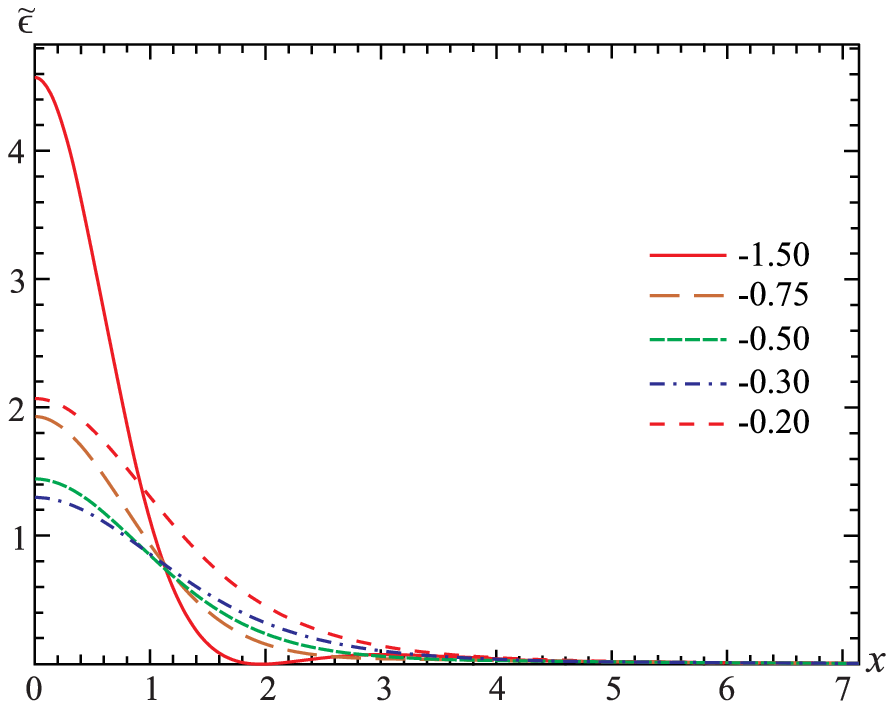}%
	}\hfill
	\subfloat[ $\tilde \xi_0=0.5, f_2 = - 0.95$.
	\label{enVs_x_E}]{
  	\includegraphics[width=0.45\linewidth]{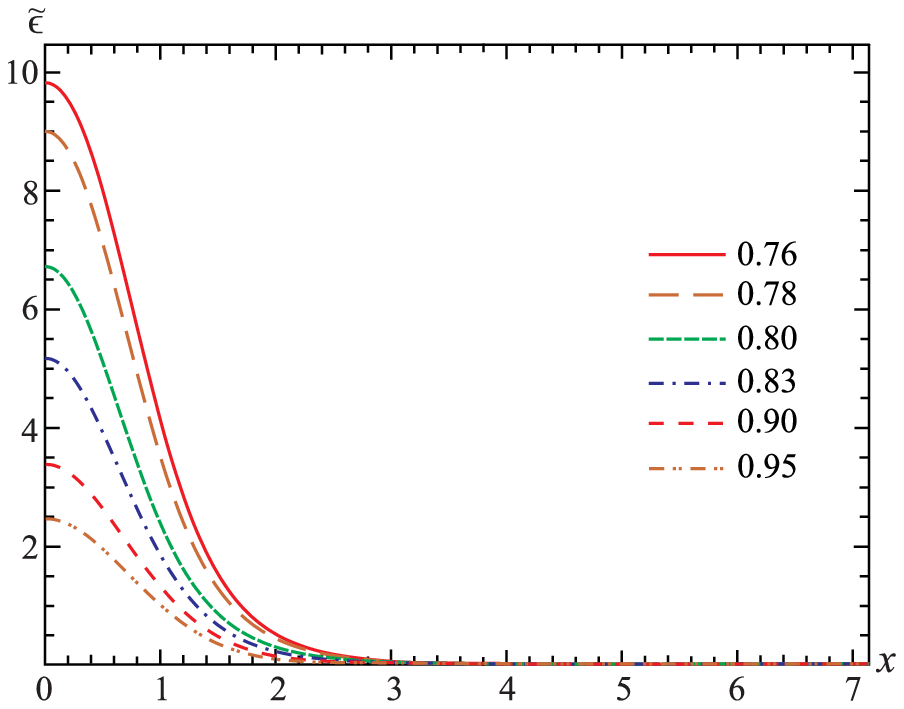}%
	}\hfill
	\hfill
	\subfloat[ $f_2=-0.6, \tilde E = 0.96$.
	\label{enVs_x_xi0}]{
  	\includegraphics[width=0.45\linewidth]{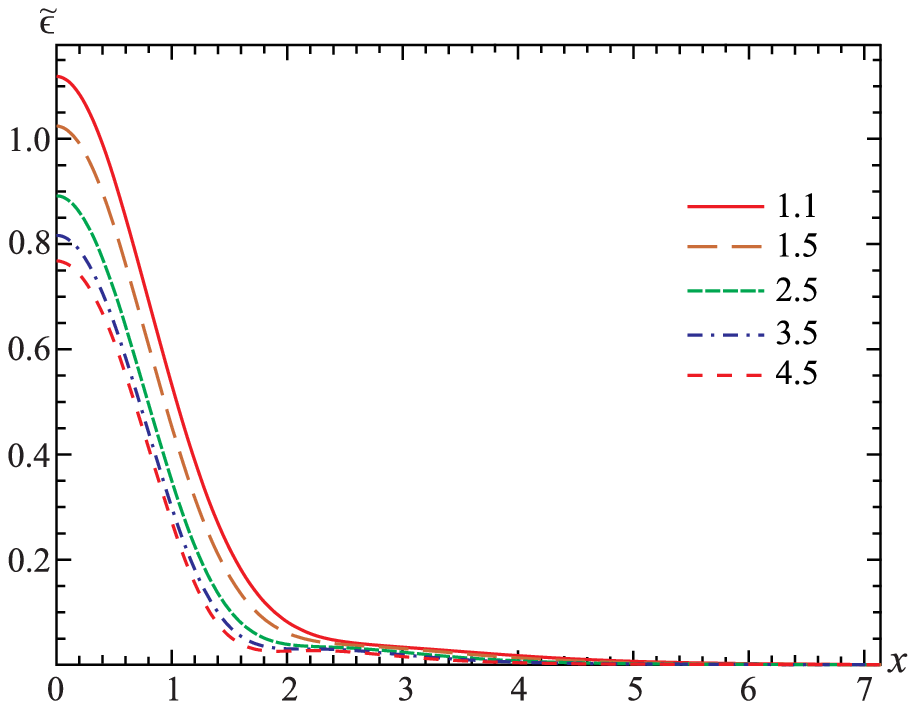}%
	}
\caption{
The energy density $\tilde \epsilon(x)$ of the solution describing the Proca-monopole-plus-spinball system with	$\tilde \Lambda = 1/9$, $\tilde m_f = 1$, $m = 3$, $\tilde g = 1$, $\tilde \lambda = 0.1$
for different values of $f_2$ [Subfig.~(a)], $\tilde E$  [Subfig.~(b)], and $\tilde \xi_0$ [Subfig.~(c)]. The corresponding values of $f_2, \tilde E$, and $\tilde \xi_0$ are designated by the numbers near the curves.
}

\label{en_dens}
\end{figure}

The dimensionless energy density of the system in question can be obtained from Eq.~\eqref{2-130-b}.
Fig.~\ref{en_dens} depicts distributions of the energy density for different $f_2$  (Fig. \ref{enVs_x_f2}), $\tilde E$ (Fig. \ref{enVs_x_E}),  and $\tilde \xi_0$  (Fig. \ref{enVs_x_xi0}).
	
\begin{figure}[h]
	\subfloat[$\tilde \xi_0=1.1, \tilde E = 0.96$.
	\label{magnProcaF_x_f2}]{
  	\includegraphics[width=0.45\linewidth]{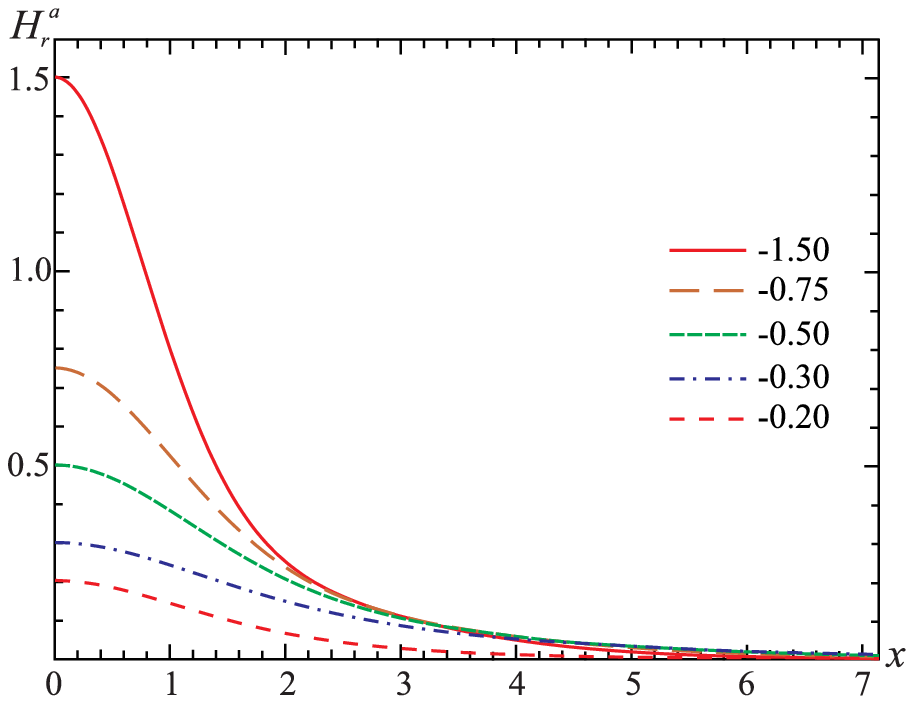}%
	}\hfill
	\subfloat[ $\tilde \xi_0=0.5, f_2 = - 0.95$.
	\label{magnProcaF_x_E}]{
  	\includegraphics[width=0.45\linewidth]{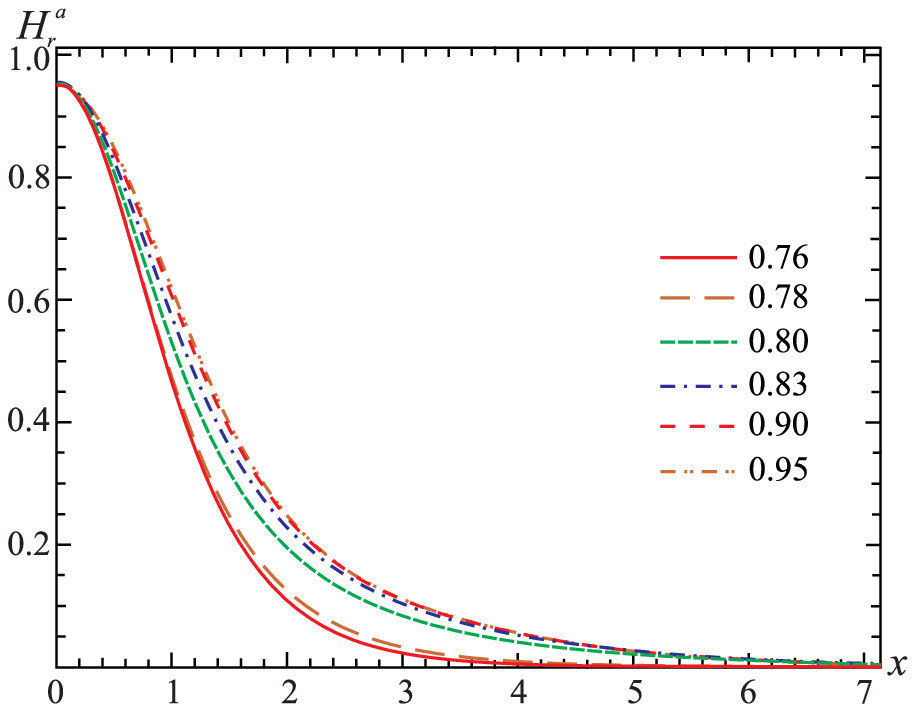}%
	}\hfill
	\hfill
	\subfloat[ $f_2=-0.6, \tilde E = 0.96$.
	\label{magnProcaF_x_xi0}]{
  	\includegraphics[width=0.45\linewidth]{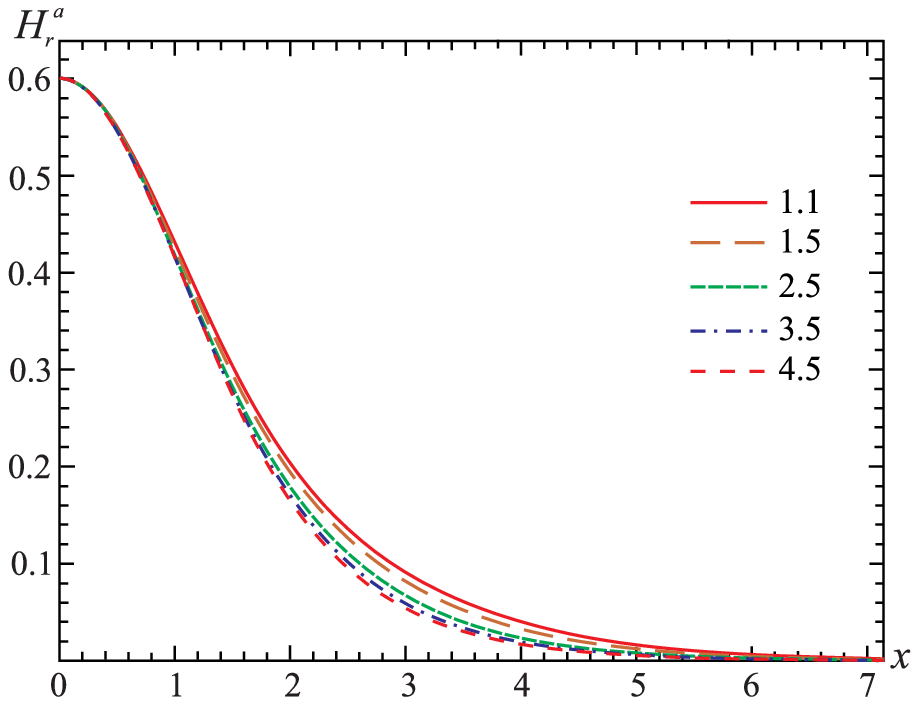}%
	}
\caption{The ``magnetic'' Proca field $H^a_r(x)$ of the solution describing the Proca-monopole-plus-spinball system with	$\tilde \Lambda = 1/9$, $\tilde m_f = 1$, $m = 3$, $\tilde g = 1$, $\tilde \lambda = 0.1$
for different values of $f_2$ [Subfig.~(a)], $\tilde E$  [Subfig.~(b)], and $\tilde \xi_0$ [Subfig.~(c)]. The corresponding values of $f_2, \tilde E$, and $\tilde \xi_0$ are designated by the numbers near the curves.
}
\label{magnProcaF}
\end{figure}

It is of great interest to follow the behaviour of the ``magnetic'' Proca field (we have been using the word ``magnetic''  in quotation marks because
the concept of a magnetic  field is not quite correct as applied to a Proca field). The radial color ``magnetic'' field is defined as follows:
\begin{equation}
		H^a_r = \frac{1 - f^2}{g r^2}.
\label{magn_field}
\end{equation}
The presence of this field enables us to speak of the Proca monopole. Its asymptotic behaviour is
\begin{equation}
	H^a_r(r) \approx \frac{2}{g}
	\frac{e^{-\frac{r}{r_0} \sqrt{m^2 \tilde M^2 - \tilde \mu^2}}}{r^2}.
\label{magn_field_asympt}
\end{equation}
It is seen from this expression that the Proca monopole differs in principle from the 't~Hooft-Polyakov monopole by its asymptotic behaviour. The graphs for the radial ``magnetic'' Proca field $H^a_r$ are given in Fig.~\ref{magnProcaF}.

\section{
Energy spectrum of the particle-like solutions for fixed $\tilde \xi_0$
}
\label{MassGap}

In this section we study the energy spectrum of the particle-like solutions as a function of the tree parameters $f_2, \tilde E$, and $\tilde \xi_0$.
Each such solution describes a ball consisting of the Dirac, Proca, and Higgs fields. Its structure is described by Eqs.~\eqref{2-50}-\eqref{2-80},
using which we have constructed the corresponding regular solutions in Sec.~\ref{QMS}. Here, we find the energy spectrum of such a system and demonstrate the presence of a mass gap for fixed $\tilde \xi_0$.

Using the expression for the dimensionless energy density $\tilde \epsilon$ from Eq.~\eqref{2-130-b}, the dimensionless total energy of the system in question is calculated as
 \begin{equation}
	\tilde W_t \equiv \frac{W_t}{ \hbar c/r_0  } = 4 \pi
	\int\limits_0^\infty x^2 \tilde \epsilon d x =
	\left( \tilde{W}_t \right)_{\text{Pm}} + \left( \tilde{W}_t \right)_{s}.
\label{4-b-10}
\end{equation}
Here, according to the decomposition \eqref{2-130-b}, we have separated the Proca monopole and spinball energies.
Solving the set of equations \eqref{2-50}-\eqref{2-80} numerically, we have computed this energy for different values of $f_2, \tilde E$, and $\tilde \xi_0$.
Our strategy of studying the energy spectrum is as follows: (i)~We obtain energy spectra for different values of $\tilde \xi_0$.
(ii)~For the spectra obtained, we show the presence of minimum of the energy, i.e., of the mass gap  $\Delta$ for a given $\tilde \xi_0$,  $\Delta(\tilde \xi_0)$.
(iii)~We examine a behaviour of $\Delta$ as a function of $\tilde \xi_0$.

\begin{figure}[h]
	\subfloat[3D plot of the energy.
	\label{enVs_f2_E_xi_08_CP}]{
  	\includegraphics[width=0.48\linewidth]{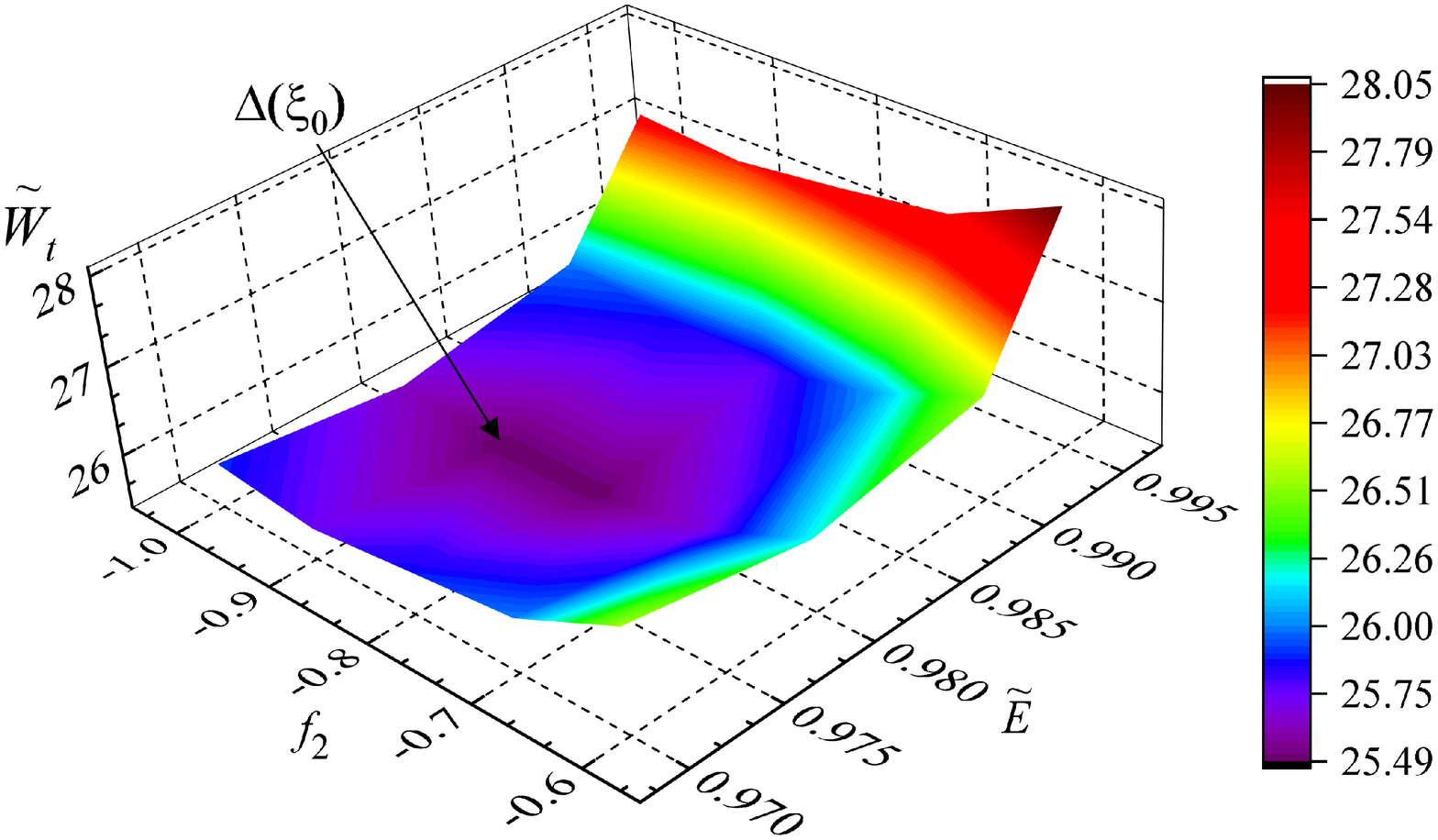}
	}\hfill
	\subfloat[Contour plot of the energy.
	\label{enVs_f2_E_xi_08_3D}]{
  	\includegraphics[width=0.48\linewidth]{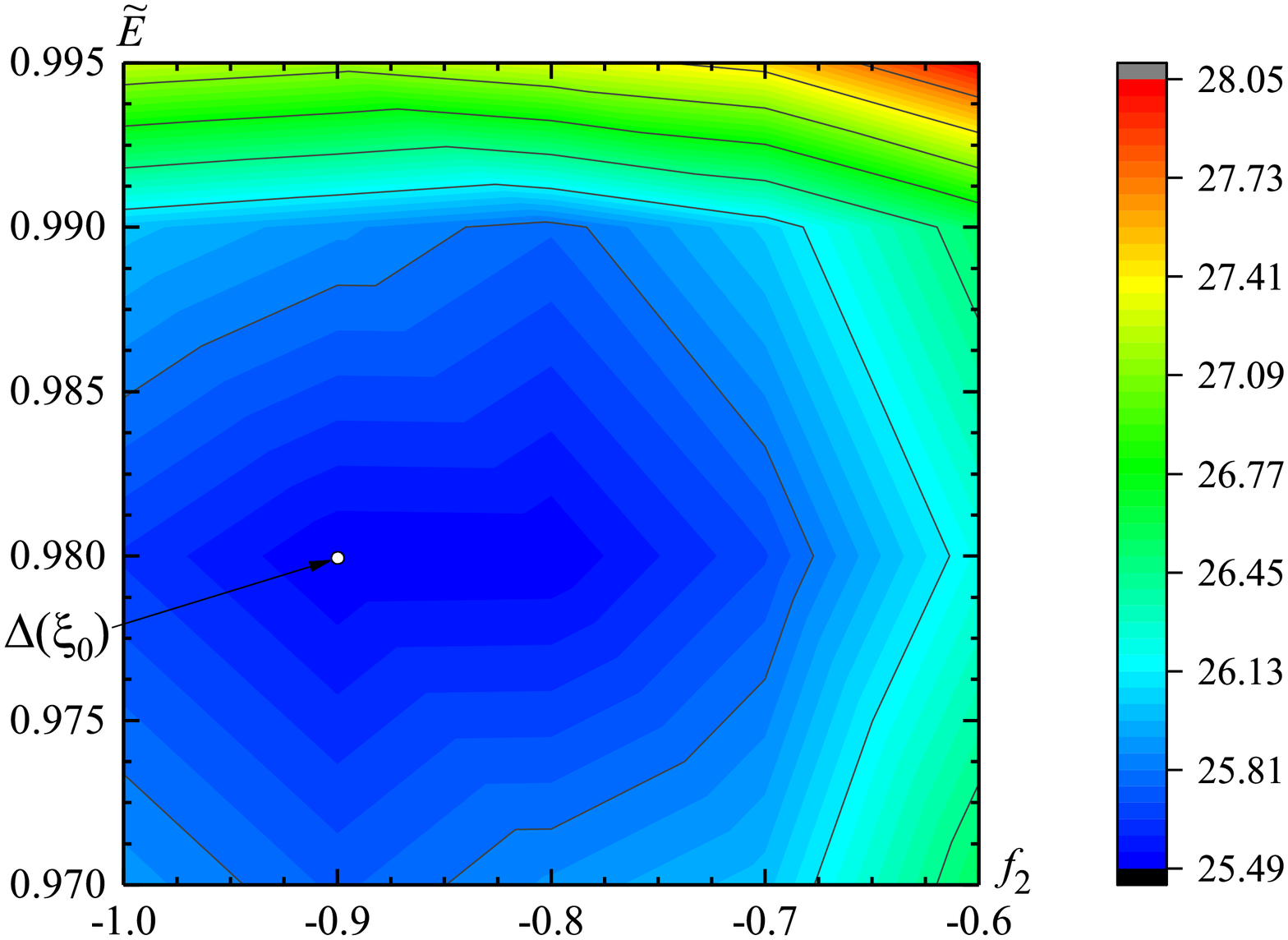}
	}
\caption{
	3D and contour plots of the energy $\tilde W_t$ from \eqref{4-b-10}
for $\tilde \xi_0 = 0.8$ as a function of $f_2$ and $\tilde E$.
}
\label{enVs_f2_E_xi_08}
\end{figure}

\begin{figure}[h]
	\subfloat[3D plot of the energy.
	\label{en_vs_f2_E_xi0_05_3D}]{
  	\includegraphics[width=0.45\linewidth]{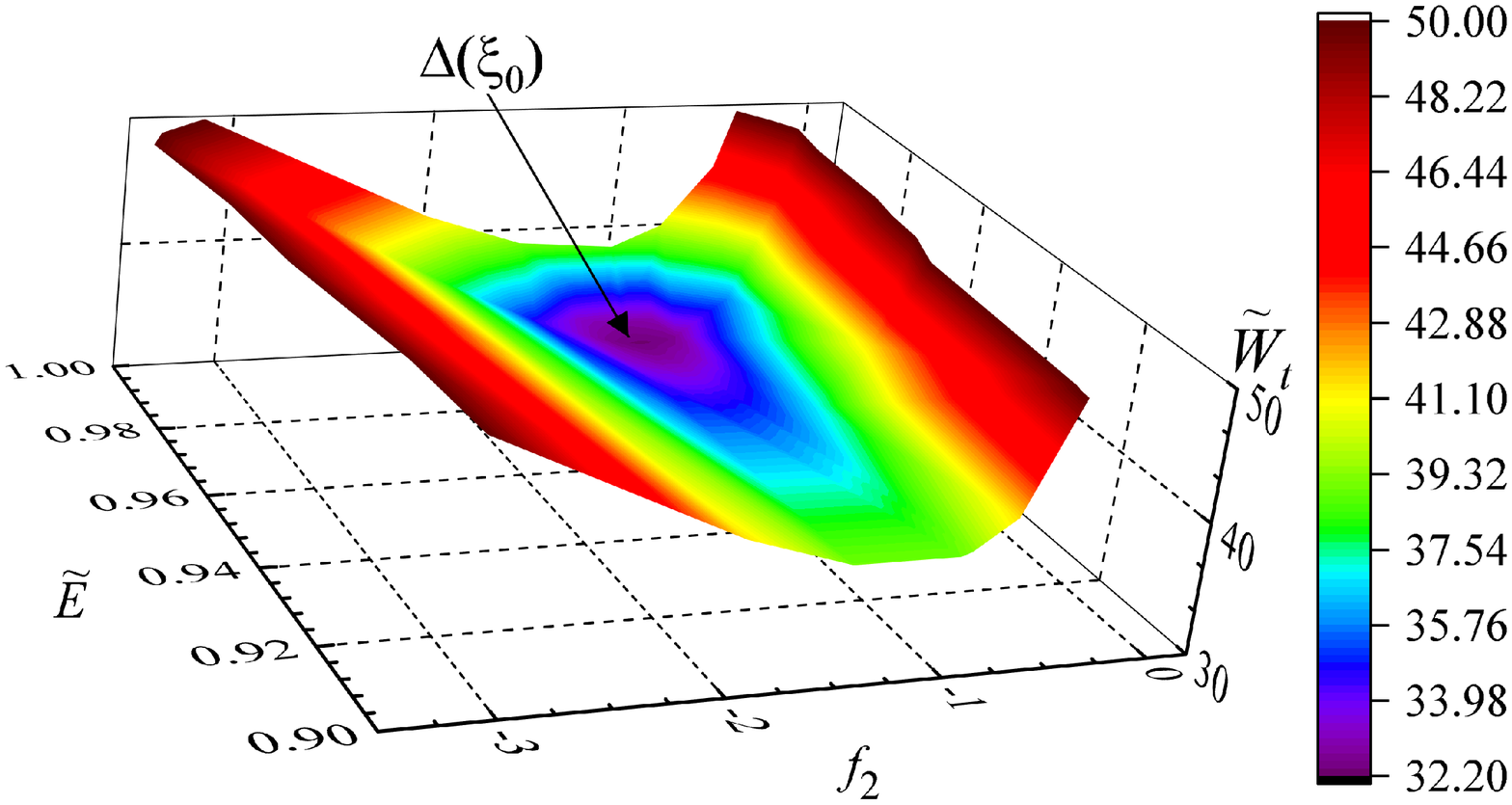}%
	}\hfill
	\subfloat[Contour plot of the energy.
	\label{en_vs_f2_E_xi0_05_CP}]{%
  	\includegraphics[width=0.45\linewidth]{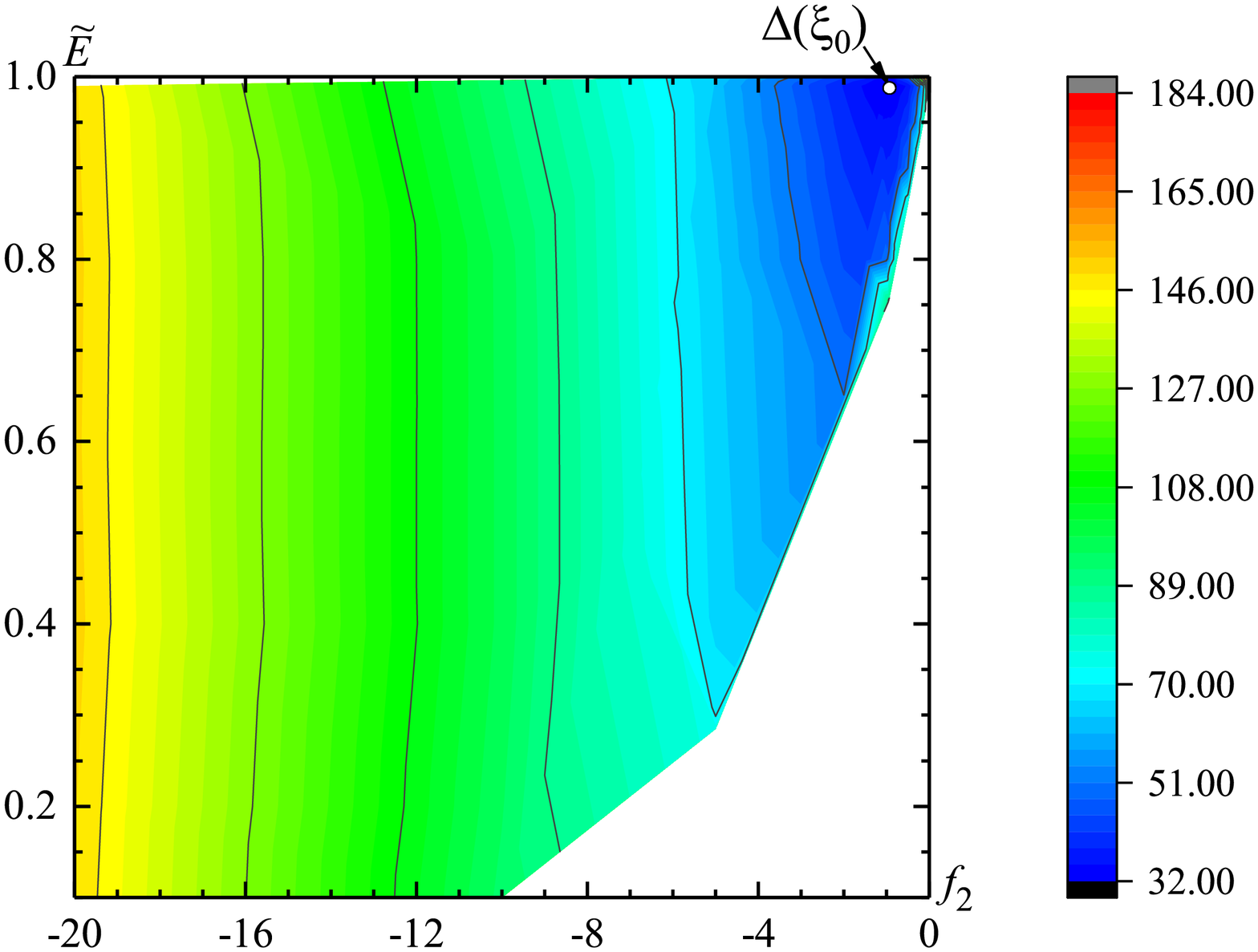}%
	}
\caption{
	3D and contour plots of the energy $\tilde W_t$ from \eqref{4-b-10}
for $\tilde \xi_0 = 0.5$ as a function of $f_2$ and $\tilde E$.
}
\label{en_vs_f2_E_xi0_05}
\end{figure}

Three-dimensional and contour plots for the energy \eqref{4-b-10} are given in Fig.~\ref{enVs_f2_E_xi_08}. It is seen from Fig.~\ref{enVs_f2_E_xi_08_3D}
that there are closed lines characterising the presence of extremum of the energy. In the case under consideration, this is a minimum corresponding to the mass gap $\Delta(\tilde \xi_0)$ for a given $\tilde \xi_0$.
The approximate value of the dimensionless mass gap for the values of the parameters used here
($\tilde \Lambda = 1/9,\tilde  m_f = 1, \tilde g = 1$, and $m = 3$) is
\begin{equation}
	\tilde \Delta(\tilde \xi_0 = 0.8)\equiv
	 \frac{\Delta(\tilde \xi_0 = 0.8)}{\hbar c/r_0} \approx 25.5 \quad
	\text{ for } \quad
	f_2 \approx -0.86, \; \tilde E \approx 0.98~.
\label{4-b-20}
\end{equation}
(For simplicity, below we omit the tilde by $\Delta$.)
Similar graphs for the energy of the particle-like solutions can be obtained for other values of $\tilde \xi_0$ as well.
In particular, Fig.~\ref{en_vs_f2_E_xi0_05} illustrates the behaviour of the energy  \eqref{4-b-10} in a wide range of values of the parameters
 $f_2$ and $\tilde E$ for $\tilde \xi_0 = 0.5$.

The numerical computations, as well as Fig.~\ref{enVs_f2_E_xi_08},  indicate that for any fixed $\tilde \xi_0$ (at least in the range
$0.4 < \tilde \xi_0 < 4.5$ under investigation) there is a minimum value of the energy, which can be called a mass gap $\Delta(\tilde \xi_0)$ for a fixed value of $\tilde \xi_0$.

Notice here the following important features of the system under consideration:
\begin{itemize}
\item
	The particle-like solutions obtained describe objects consisting of a Proca monopole and of a spinball supported by strongly interacting non-Abelian SU(2) Proca vector fields and nonlinear spinor and scalar fields.
\item
	The numerical calculations enable us to speak with certainty that for any fixed value of  $\tilde \xi_0$ there is a minimum value of the energy, which can be called a mass gap $\Delta(\tilde \xi_0)$ for fixed $\tilde \xi_0$,
i.e., there is a dependence $\Delta(\tilde \xi_0)$.
\item
	The numerical analysis does not allow us to determine whether there is a minimum value of the energy 	(a global value of the mass gap $\Delta$) in a whole range of possible values of the parameter $\tilde \xi_0$.
	The analysis of the behaviour of $\Delta(\tilde \xi_0)$ in the range
	$0.4 < \tilde \xi_0 < 4.5$ permits us to assume that
	$
		\Delta(\tilde \xi_0) \xrightarrow{\tilde \xi_0 \rightarrow \infty}
		\begin{cases}
			\text{either const}\\
			\text{or} \quad \quad \infty
		\end{cases}
	$.
\item
	The numerical analysis indicates that as $\tilde E \rightarrow 1$, the total energy $\tilde W_t \rightarrow + \infty$.
\item
	Regular solutions of Eqs.~\eqref{2-50}-\eqref{2-80} exist not for all pairs $\{ f_2, \tilde E \}$.
  It is seen from Fig.~\ref{en_vs_f2_E_xi0_05} that for some fixed  $f_2$ there exists a critical value $\tilde E_{\text{cr}}$
 for which the solution still exists but no solutions are found for $\tilde E < \tilde E_{\text{cr}}$.
  According to the numerical calculations, the energy of the solutions obtained tends to infinity when
   $\tilde E\rightarrow\tilde E_{\text{cr}}$, i.e., $\tilde W_t \xrightarrow{\tilde E\rightarrow\tilde E_{\text{cr}}} + \infty$.
\end{itemize}

\section{Analysis of the behaviour of the mass gap $\Delta(\tilde \xi_0)$ for different $\tilde \xi_0$}

In Sec.~\ref{QMS}, we have shown that the set of equations describing the Proca-Dirac-Higgs system has particle-like
 Proca-monopole-plus-spinball solutions.
In Sec.~\ref{MassGap},
we have studied the energy spectrum of the solutions for fixed  values of the parameter $\tilde \xi_0$ and have shown that the spectrum possesses a mass gap
 $\Delta(\tilde \xi_0)$ for some values of $\tilde \xi_0$.

The next step is to study the behaviour of the mass gap  $\Delta(\tilde \xi_0)$ in  a whole range of values of the parameter $\tilde \xi_0$.
The significance of this problem is due to the fact that if the quantity  $\Delta(\tilde \xi_0)$ has a global minimum this means that the Proca-Dirac-Higgs theory has a mass gap.
The problem of presence or absence of a mass gap is of great importance in modern quantum field theory. Its solving would practically mean solving the problem of a nonpertubative quantization,
and this is necessary, for instance, in constructing a theory of strong interactions.
In the present paper we study a classical field theory based on the coupled set of the Proca, nonlinear Klein-Gordon and Dirac equations.
It seems to us that the presence of a mass gap, even in a classical field theory,  would be of great interest since it would allow one to get insight into the reason for its appearance.

\begin{figure}[t]
\begin{minipage}[t]{.45\linewidth}
	\begin{center}
		\includegraphics[width=1\linewidth]{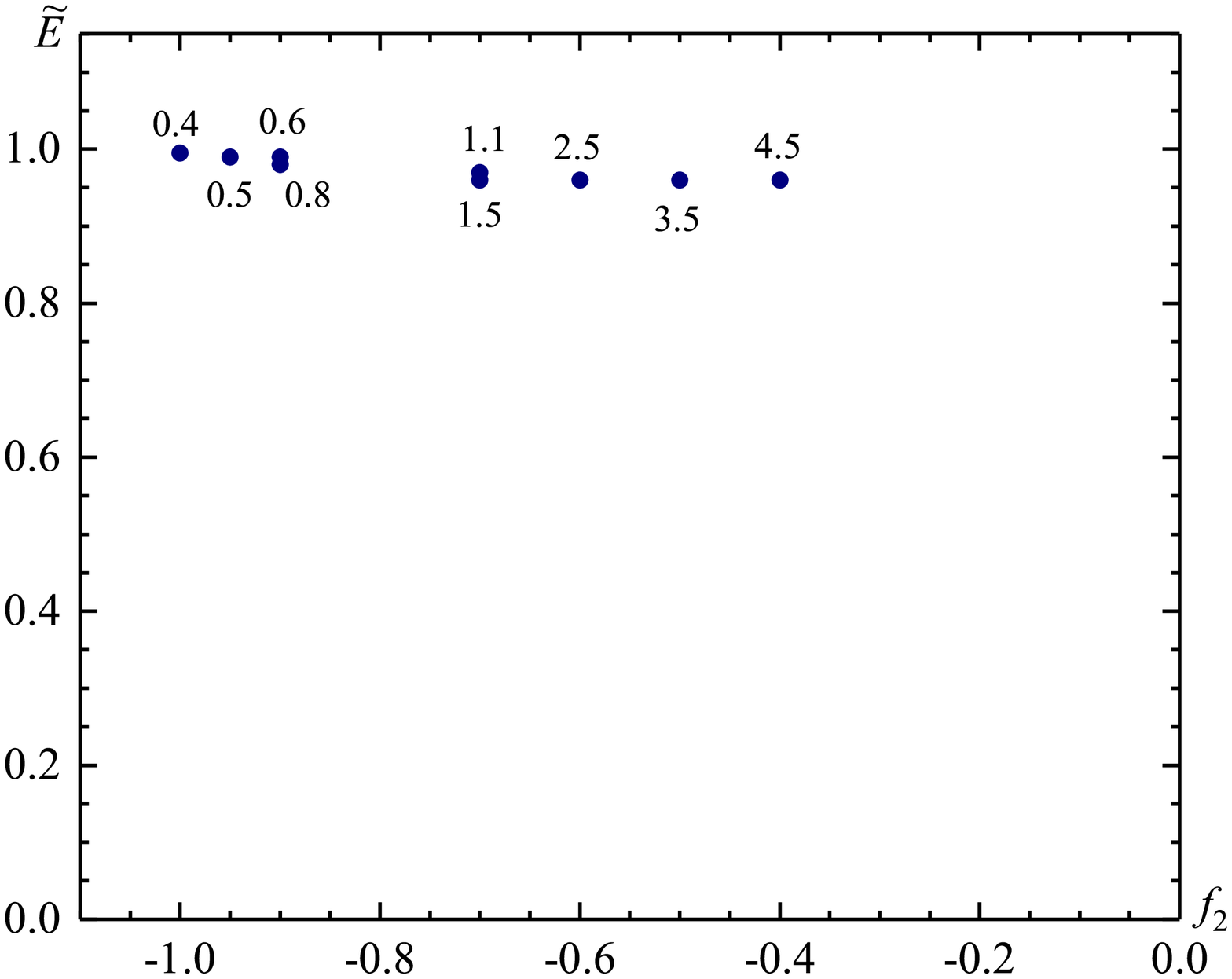}
	\end{center}
	\vspace{-0.5cm}
	\caption{The positions of minima of the energy $\tilde W_t$.
	The numbers near the points denote the corresponding values of $\tilde \xi_0$.
}
	\label{locations_MGs}
\end{minipage}\hfill
\begin{minipage}[t]{.45\linewidth}
	\begin{center}
		\includegraphics[width=1\linewidth]{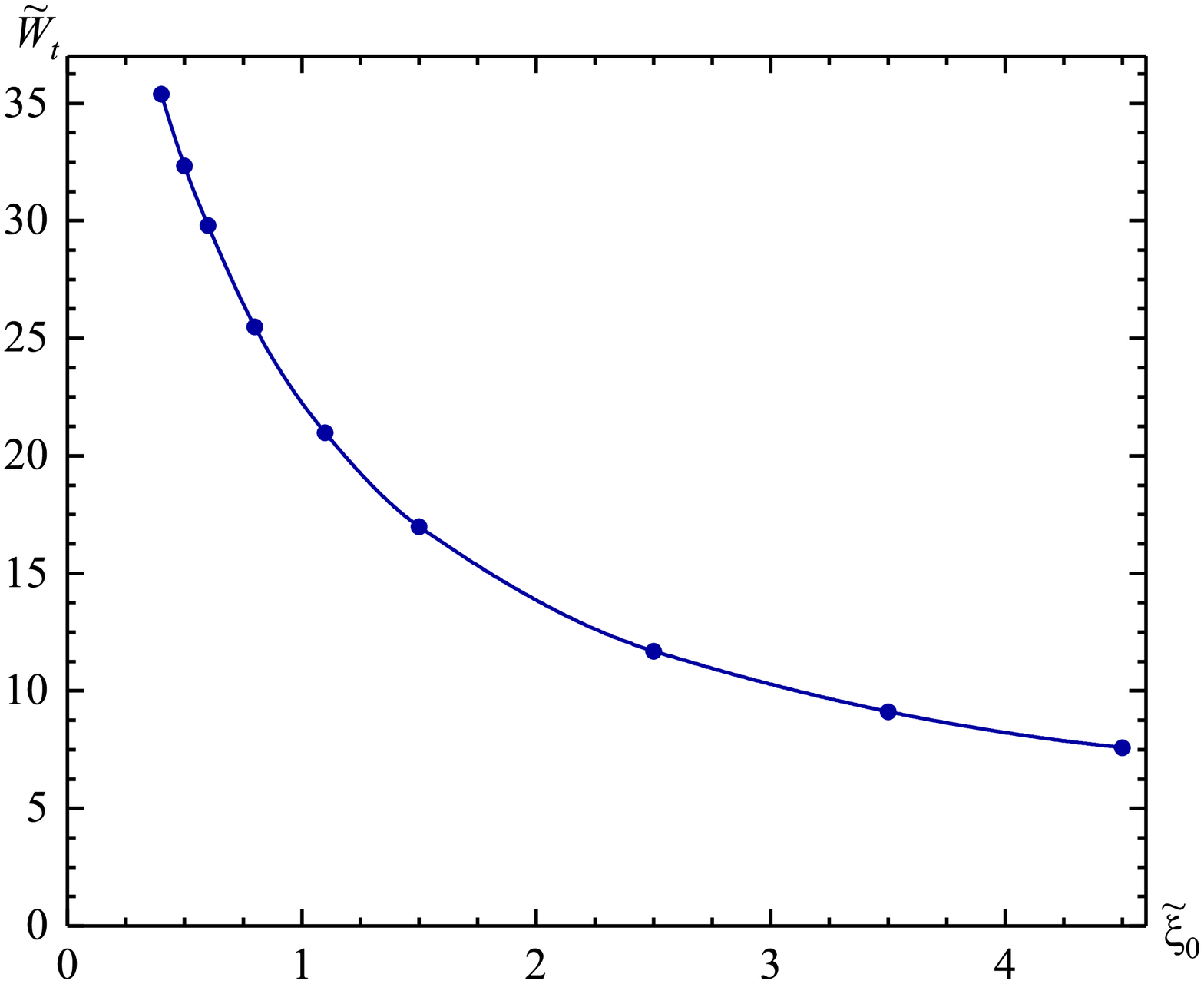}
	\end{center}
	\vspace{-0.5cm}
	\caption{
		The minima of the energy $\tilde W_t$ as a function of $\tilde \xi_0$.
	}
	\label{Delta_xi0}
\end{minipage}\hfill
\end{figure}

To study the question of the presence of the mass gap in a whole range of values of the parameters $f_2, \tilde E$, and $\tilde \xi_0$, it is necessary to calculate   $\Delta(\tilde \xi_0)$ for
$0 < \tilde \xi_0 < \infty$ and to check whether or not there is a global minimum of this function. Fig.~\ref{Delta_xi0} shows this curve in the range $0.4 < \tilde \xi_0 < 4.5$. Unfortunately, deriving solutions for small $f_2$ and large $\tilde \xi_0$ runs into great technical difficulty: it is impossible to find eigenvalues of $\tilde \mu, \tilde M$, and $u_1$ to the necessary accuracy.

The positions of minima of the energy \eqref{4-b-10} on the plane
$\{f_2, \tilde E\}$ are illustrated by Fig.~\ref{locations_MGs} for different values of $\tilde \xi_0$. Notice that numerical errors in calculating magnitudes and positions of minima of the energy $\Delta(\tilde \xi_0)$
are quite large, and perhaps this leads to  shifting the points with $\Delta(\tilde \xi_0 = 0.6, 0.8)$ and $\Delta(\tilde \xi_0 = 1.1, 1.5)$.

\begin{figure}[t]
\centering
		\includegraphics[width=.45\linewidth]{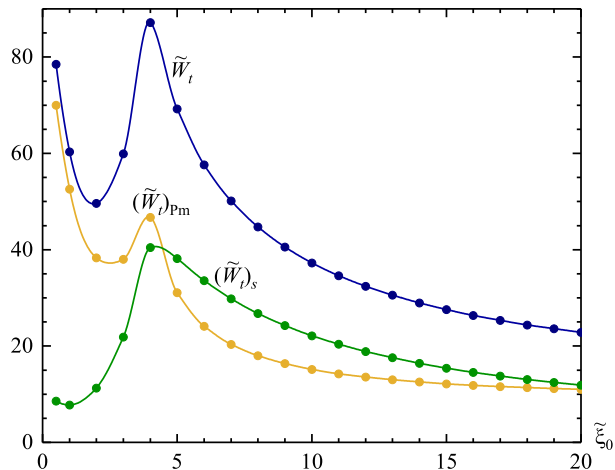}
	
	\caption{The energies $\tilde W_t$, $\left( \tilde{W}_t \right)_{\text{Pm}}$, and $\left( \tilde{W}_t \right)_{s}$ as functions
of the parameter $\tilde \xi_0$ for $f_2 = -7$ and $\tilde{E} = 0.5$.
}
	\label{fig_W_vs_xi0_f2_7_E_0_5}
\end{figure}

Search for the mass gap $\Delta(\tilde \xi_0)$ for $\tilde \xi_0 > 4.5$ runs into the aforementioned technical problems. For a qualitative understanding what can happen for large $\tilde \xi_0$,
we have investigated dependencies of the total energy, $\tilde{W}_t$, and of the energies of the monopole,
$\left( \tilde{W}_t \right)_{\text{Pm}}$, and of the spinball, $\left( \tilde{W}_t \right)_{s}$, on the parameter $\tilde \xi_0$
for fixed values of the parameters $f_2$ and $\tilde{E}$. Taking into account the formulas \eqref{2-120} and \eqref{2-130-c},
the quantities $\left( \tilde{W}_t \right)_{s}$ and $\left( \tilde{W}_t \right)_{\text{Pm}}$ describe the total energy of the system under investigation as a sum of the energies of the Proca monopole and of the spinball:
$
	\tilde{W}_t = \left( \tilde{W}_t \right)_{\text{Pm}} +
	\left( \tilde{W}_t \right)_{s}
$. The results of calculations are given in Fig.~\ref{fig_W_vs_xi0_f2_7_E_0_5}. It is seen that there are local minima for all the energies for some values of $\tilde \xi_0$.
The behaviour of these quantities when the parameter $\tilde \xi_0$ increases further is unclear, and it requires further investigations.
From an analysis of the graphs, we can assume that
$
	\left( \tilde{W}_t \right)_{s}
$ will tend to zero and
$
	\left( \tilde{W}_t \right)_{\text{Pm}}
$ will tend to a constant value as $\tilde \xi_0 \to\infty$; this would mean that
$
	\tilde{W}_t	\xrightarrow{\tilde \xi_0 \rightarrow \infty} \mathrm{const}
$
for fixed $f_2$ and $\tilde{E}$.
This observation permits us to assume that such a behaviour persists for all values of  $f_2$ and $\tilde{E}$;
this may result in the existence of a mass gap in the energy spectrum of the particle-like solutions under investigation within a classical field theory containing a non-Abelian Proca field plus nonlinear Higgs and
Dirac fields.

\begin{figure}[!htb]
	\subfloat[The function $f(x)$.
	\label{fig_f_vs_x_f2_7_E_05_var_xi0}]{
  	\includegraphics[width=0.45\linewidth]{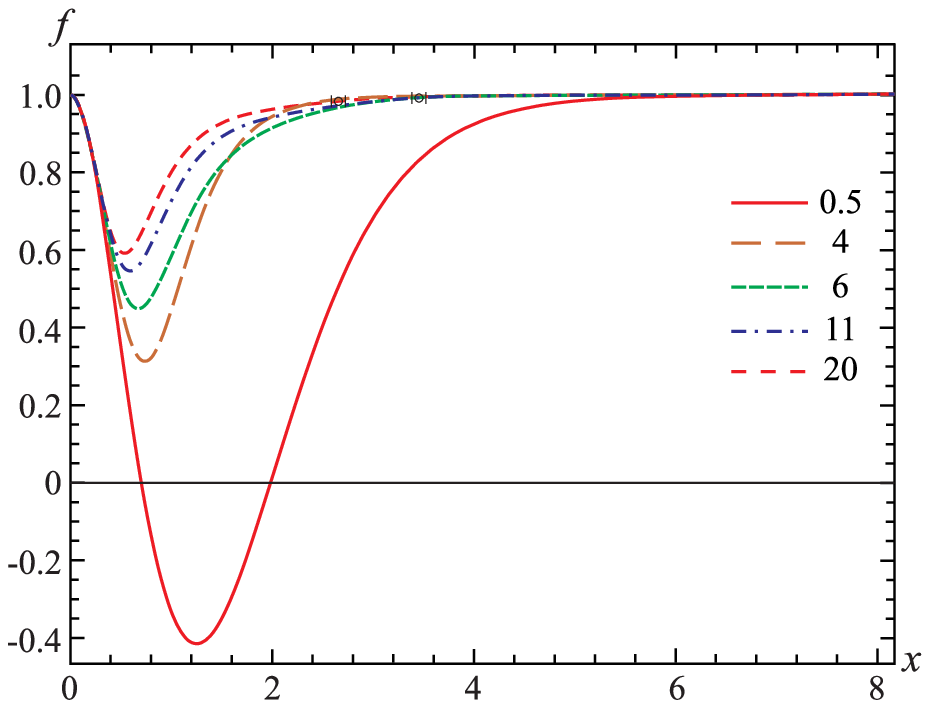}%
	}\hfill
	\subfloat[The  function $\tilde \xi(x) - \tilde \xi_0$.
	\label{fig_xi_vs_x_f2_7_E_05_var_xi0}]{%
  	\includegraphics[width=0.45\linewidth]{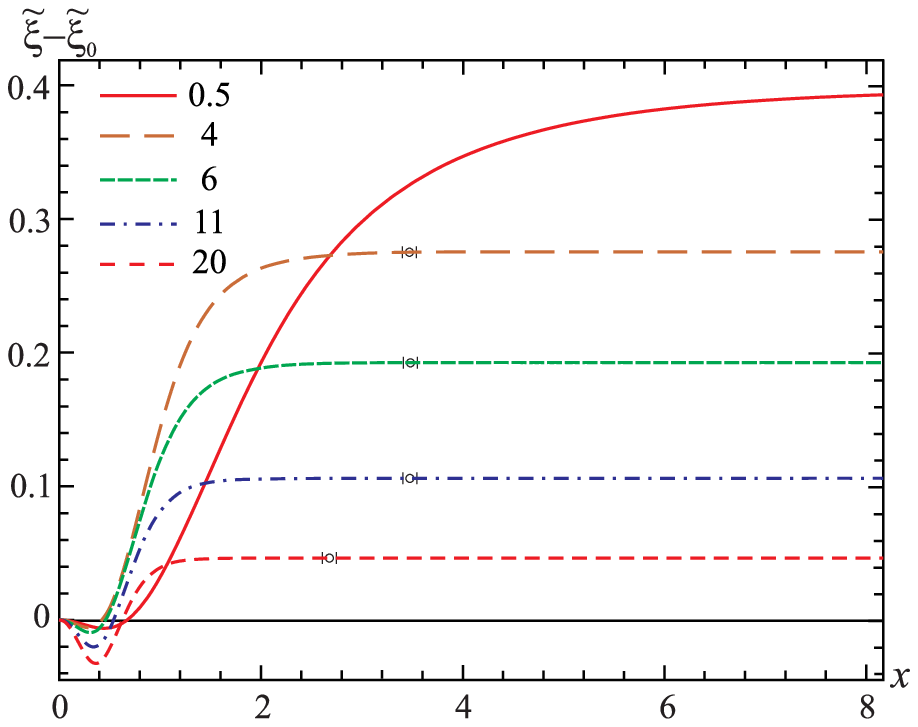}%
	}\hfill
	\subfloat[The function $\tilde u(x)/x$.
	\label{fig_u_vs_x_f2_7_E_05_var_xi0}]{
  	\includegraphics[width=0.45\linewidth]{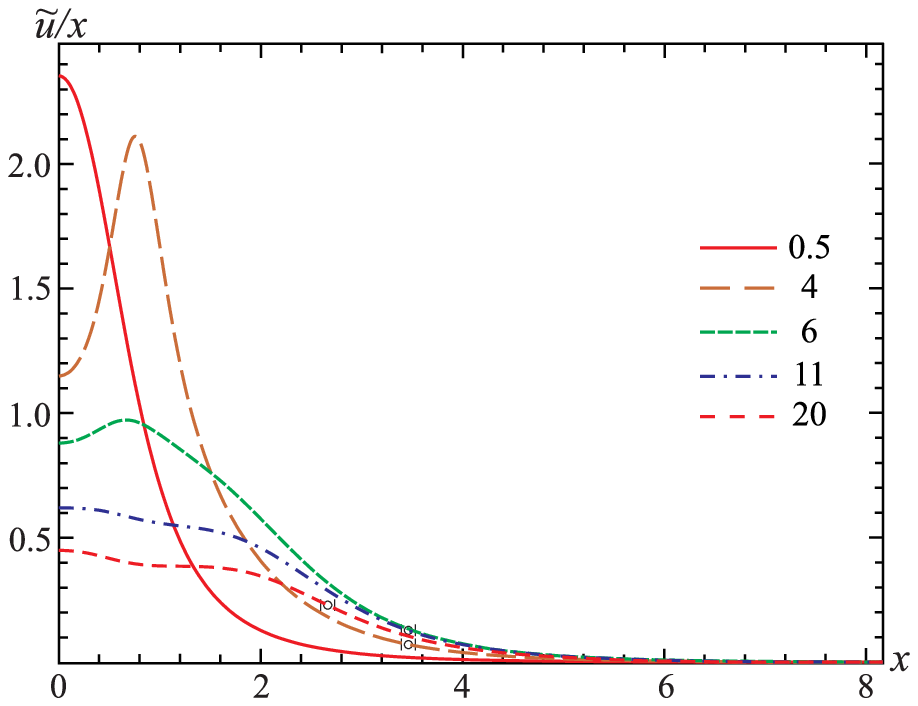}%
	}\hfill
	\subfloat[The function $\tilde v/x$.
	\label{fig_v_vs_x_f2_7_E_05_var_xi0}]{%
  	\includegraphics[width=0.45\linewidth]{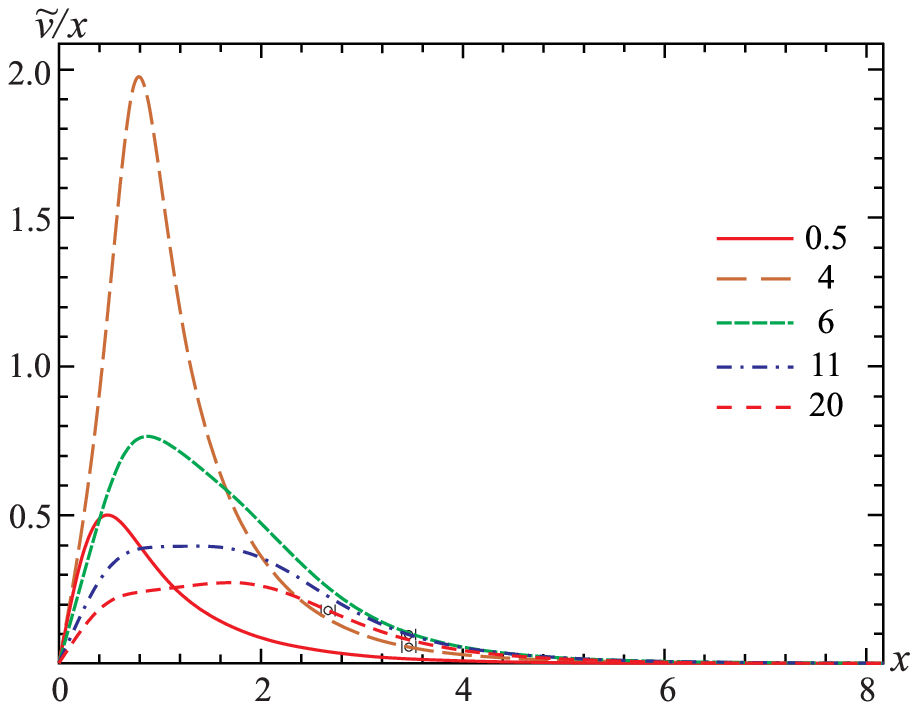}%
	}
\caption{
	Families of particle-like solutions for $f_2 = - 7$ and $\tilde E = 0.5$ for the Proca-monopole-plus-spinball system with
	$\tilde \Lambda = 1/9$, $\tilde m_f = 1$, $m = 3$, $\tilde g = 1$, $\tilde \lambda = 0.1$ for different values of the parameter $\tilde \xi_0$ (designated by the numbers near the curves).
The symbols  $| \! \! \circ \! \! |$ mark the points where (i) in Subfigs.~(a) and (b), the numerical solutions for the functions $f$ and $\tilde \xi$
match with the asymptotic solutions \eqref{3-c-90} and \eqref{3_c_120}; (ii) in Subfigs.~(c) and (d),
for the functions $\tilde u$ and $\tilde v$, we sought a numerical solution of only two equations \eqref{2-70} and \eqref{2-80}
with replacing the functions $f$ and $\tilde \xi$ by the asymptotic expressions \eqref{3-c-90} and \eqref{3_c_120}.
}
\label{fig_f_xi_u_v_vs_x_f2_7_E_05_var_xi0}
\end{figure}

For understanding the behaviour of the energies
$
	\tilde{W}_t, \left( \tilde{W}_t \right)_{\text{Pm}}$, and
	$\left( \tilde{W}_t \right)_{s}$
as functions of $\tilde \xi_0$, we show in Fig.~\ref{fig_f_xi_u_v_vs_x_f2_7_E_05_var_xi0} the graphs of the functions
$f(x), \tilde \xi(x), \tilde u(x)/x$, and $\tilde v(x)/x$.
It is seen that with increasing $\tilde \xi_0$ characteristic magnitudes of the functions $f$ and  $\tilde \xi$ decrease, while those of the functions $\tilde u/x$ and $\tilde v/x$
increase. Such a behaviour of the functions leads at least to not decreasing (and maybe even to increasing) the energy of the Proca monopole, which basically depends on the functions $f$ and $\tilde \xi$,
and to increasing the energy of the  spinball, which basically depends on the functions  $\tilde v$ and $\tilde u$.
For large values of $\tilde \xi_0$, the solutions are derived by matching numerical solutions (which are obtained up to some radius $x_1$)
with the corresponding asymptotic functions from \eqref{3-c-90} and \eqref{3_c_120}.

\section{Discussion and conclusions}
\label{concl}

We have considered non-Abelian SU(2) Proca-Dirac-Higgs theory where a massive vector field interacts with nonlinear scalar and spinor fields.
Within such theory, regular, nontrivial particle-like solutions describing finite-energy configurations~--
 Proca-monopole-plus-spinball systems~-- have been constructed. A distinctive feature of the Proca monopole is that, unlike the 't~Hooft-Polyakov monopole, the corresponding color field decays exponentially at infinity. In our opinion, the main reason why such solutions can exist is the presence of the mass of the non-Abelian Proca field,
and also because of the structure of the Dirac equation leading to the existence of regular solutions of this equation.

Note that both a Proca monopole and a spinball can exist separately, i.e., one can get a Proca monopole as a solution of Eqs.~\eqref{2-50}  and \eqref{2-60} with $\tilde u = \tilde v = 0$.
Also, there exists a particle-like solution (a spinball) of the nonlinear Dirac equation
 \eqref{2-70} and \eqref{2-80} with
$f = 1$ and $\xi = \mathrm{const}$ (see Refs.~\cite{Finkelstein:1951zz,Finkelstein:1956}).
The basic distinction between these two solutions is that the energy spectrum of the Proca monopole does not have a mass gap,
while for the particle-like solution of the nonlinear Dirac equation such a gap does exist~\cite{Finkelstein:1951zz,Finkelstein:1956}.
This means that if the mass gap exists for the complete set of equations \eqref{2-50}-\eqref{2-80},  this is a consequence of the presence of the nonlinear Dirac field.
Note in this connection that in the 1950's the mass gap has in fact been found in Refs.~\cite{Finkelstein:1951zz,Finkelstein:1956}
in solving the nonlinear Dirac equation. However, the authors did not use such term, but said of ``the lightest stable particle''.
Those papers were devoted to study of the nonlinear Dirac equation, and  W.~Heisenberg offered to use it as a fundamental equation in describing the properties of an electron.
To the best of our knowledge, the mass gap was first obtained in Refs.~\cite{Finkelstein:1951zz,Finkelstein:1956}.

We have calculated the energy spectrum of the corresponding solutions for fixed values of $\tilde \xi_0$ and have shown that there is a minimum value of the energy
(the mass gap $\Delta(\tilde \xi_0)$ for a fixed $\tilde \xi_0$). Also, we have studied the dependence of $\Delta$ on $\tilde \xi_0$ in the range $0.4 < \tilde \xi_0 < 4.5$. In order to understand whether the mass gap exists in a whole range of possible values of $\tilde \xi_0$, we have examined the behaviour of the total energy of the  particle-like solution, as well as the behaviour of the
energies of the Proca monopole and of the spinball, which make up the particle-like configuration under consideration, for some fixed values of the parameters  $f_2$ and $\tilde E$. As a result, we have shown that these energies have local minima, and the energy of the Proca monopole possibly tends to a constant value. This permits us to suggest that such a behaviour will occur for any values of  $f_2$ and $\tilde E$; this eventually will lead to the appearance of a mass gap in a whole range of values of the parameters $f_2, \tilde E$, and $\tilde \xi_0$ determining the characteristics of the particle-like solutions under investigation.

Thus, one of the purposes of this paper is to study  the energy spectrum of particle-like solutions and to obtain the mass gap $\Delta(\xi_0)$
for a fixed $\xi_0$. All this permits us to understand the mechanism of the occurrence of a mass gap in Proca-Dirac-Higgs theory: the reason is that the Dirac equation is nonlinear. Bearing this in mind, one can assume that a similar mechanism may also be responsible for the appearance of a mass gap in QCD. For this, however, one has to understand  how the nonlinear Dirac equation may occur in QCD. This can happen as follows. In the Lagrangian, the interaction between quarks and gluons is described by the term
$
	\hat{\bar \psi} \lambda^B \hat{A}_\mu \hat{\psi}
$, where
$
\hat{\psi} = \left\langle \hat{\psi} \right\rangle + \widehat{\delta \psi},
\hat A_\mu = \left\langle \hat{A}_\mu \right\rangle + \widehat{\delta A}_\mu
$,
$
	\left\langle \hat{\psi} \right\rangle,
	\left\langle \hat{A}_\mu \right\rangle
$ are valence quarks and gluons, and
$
 \widehat{\delta \psi}, \widehat{\delta A}_\mu
$ are sea quarks and gluons; $\lambda^B $ is the Gell-Mann matrices.
One can assume that the quantum average of the term
$
	\left\langle
		\widehat{\delta \bar{\psi}} \lambda^B \widehat{\delta A}_\mu
		\widehat{\delta \psi}
	\right\rangle
$
will \emph{approximately} look like
$
	\left\langle
		\widehat{\delta \bar{\psi}} \lambda^B \widehat{\delta A}_\mu
		\widehat{\delta \psi}
	\right\rangle \approx
	\phi \left(
		\bar \xi \xi
	\right)^2
$, where the field $\phi$  \emph{approximately} describes the sea gluons and the spinor field $\xi$ \emph{approximately} describes
the sea quarks. Thus, in QCD, there can occur the nonlinear Dirac equation which \emph{approximately} describes the interaction between sea quarks and gluons.
Summarizing, one can say that the following mechanism of the occurrence of a mass gap in QCD is suggested: the nonperturbative interaction between sea quarks and gluons is
\emph{approximately} described by a nonlinear spinor field, the presence of which leads in turn to the occurrence of a mass gap in QCD.

Thus, the following results have been obtained:
\begin{itemize}
\item
	Particle-like solutions of the type Proca-monopole-plus-spinball have been found in some range of values of the system parameters $f_2, \tilde E$, and $\tilde \xi_0$.
\item
	For such solutions, the energy spectra for some values of the parameter $\tilde \xi_0$ have been constructed. It was shown that they possess a minimum $\Delta(\tilde \xi_0)$.
\item
	The behaviour of $\Delta(\tilde \xi_0)$ as a function of the parameter $\tilde \xi_0$ has been studied.
\item
	It was shown that the solutions obtained give rise to a Meissner-like effect, which consists in the fact that a maximum value of the SU(2) gauge Proca field is located there where the Higgs field has a minimum.
\item
	For the Proca monopole, it was shown that the Proca color ``magnetic'' field decreases asymptotically according to an exponential law.
\item
    The non-Abelian Proca monopole obtained differs in principle from the 't~Hooft-Polyakov monopole in the sense that the Proca monopole is topologically trivial.
\item
    The main reason for the existence of the mass gap $\Delta(\tilde \xi_0)$ is due to the presence of the nonlinear Dirac field.
\item 
The mechanism of the occurrence of a mass gap in QCD has been suggested.
\end{itemize}

Note that when one tries to prove the presence (or absence) of the mass gap $\Delta$, he/she encounters great technical  difficulties
related to the fact that with increasing $\tilde \xi_0$ the mass gap $\Delta(\tilde \xi_0)$
shifts towards $f_2 \rightarrow 0$ and $\tilde E \rightarrow m_f$. But for such values of $f_2$ and $\tilde E$ the
eigenvalues $\tilde{\mu}, \tilde{M}$, and $\tilde{u}_1$ should be given to high accuracy, leading to such  technical problems.
In order to assume what can happen with the function $\Delta(\tilde \xi_0)$ with increasing  $\tilde \xi_0$, we have studied the behaviour
of the total energy $\tilde W_t$, as well as the energies of the Proca monopole, $\left( \tilde{W}_t \right)_{\text{Pm}}$, and of the spinball,
$\left( \tilde{W}_t \right)_{s}$, as functions of $\tilde \xi_0$
for some fixed values of the parameters $f_2, \tilde{E}$
and have shown that in this case all the energies have a minimum (at least a local one). This allows us to hope that such a minimum will occur for any values of the parameters $f_2$ and $\tilde{E}$; this in turn assumes that there will be a minimum value of the energy  $\tilde W_t$ in a whole range of values of the parameters $f_2, \tilde{E}$, and $\tilde \xi_0$.

Numerical study of the energy spectrum of a spinball [when one solves only Eqs.~\eqref{2-70} and \eqref{2-80}] indicates that
its energy goes to infinity when $\tilde E \rightarrow \tilde m_f$ and when $\tilde E \rightarrow  0$; this corresponds to the fact that somewhere inside this region there is
a minimum value of the energy (a mass gap of the spinball). Our numerical calculations indicate that apparently when Proca and Higgs fields are applied
the energy of our system also goes to infinity, but now when
$\tilde E\rightarrow \tilde m_f$ or when  $f_2$ tends to some critical value $f_2=(f_2)_{\text{cr}}$.

Taking into account all the above, the existence of a mass gap within a theory containing a SU(2) Proca field plus nonlinear Higgs and Dirac fields is not impossible. To confirm this, it is necessary to carry out further investigations in this direction which are connected with large technical problems associated with solving a nonlinear eigenvalue problem when eigenvalues must be determined to high accuracy.

Finally, note that the particle-like solutions obtained here can be considered as those describing a spinorial Proca glueball. In QCD, a glueball is a hypothetical particle consisting only of a gauge
Yang-Mills field in the absence of quarks. In our case a spinor field is present, but this is a nonlinear field that does not describe quarks.
Since the solutions obtained are supported by non-Abelian Proca and nonlinear Higgs and Dirac fields, we can refer to a configuration described by such solutions as a spinorial Proca glueball (or a Proca-monopole-plus-spinball system).

\section*{Acknowledgements}

This work was supported by Grant No.~BR05236730 in Fundamental Research in Natural Sciences by the Ministry of Education and Science of the Republic of Kazakhstan.
V.D. and V.F. also are grateful to the Research Group Linkage Programme of the Alexander von Humboldt Foundation for the support of this research.


\end{document}